\documentclass[aip, jcp,amsmath,amssymb,reprint]{revtex4-2}
\usepackage[utf8]{inputenc}
\usepackage{dcolumn}
\usepackage{bm}
\usepackage{graphicx}
\usepackage[utf8]{inputenc}
\usepackage[T1]{fontenc}
\usepackage{mathptmx}
\usepackage{color}
\usepackage{lineno}
\usepackage{etoolbox}
\usepackage{mathtools} 
\usepackage{silence}
\usepackage{orcidlink}
\definecolor{lightergray}{gray}{0.95}
\colorlet{darkorange}{orange!80!black}
\usepackage[normalem]{ulem}
\usepackage{physics}
\usepackage{booktabs} 
\usepackage{array} 
\usepackage[version=4]{mhchem}
\usepackage{tablefootnote} 

\makeatletter
\def\@email#1#2{%
 \endgroup
 \patchcmd{\titleblock@produce}
  {\frontmatter@RRAPformat}
  {\frontmatter@RRAPformat{\produce@RRAP{*#1\href{mailto:#2}{#2}}}\frontmatter@RRAPformat}
  {}{}
}%
\makeatother

\newcommand{\br}{\boldsymbol r} 
\newcommand{\mr}{\mathbf r} 

\newcommand{\Tau}{\mathrm{T}} 
\newcolumntype{R}{>{\raggedleft\arraybackslash}p{1.35cm}}  

\begin{document}


\title{Partitioning the electronic wave function using deep variational Monte Carlo}
\author{Mat\v{e}j Mezera \orcidlink{0009-0003-0047-488X}}
\homepage{m.mezera@fu-berlin.de}
\affiliation{FU Berlin, Department of Mathematics and Computer Science, Arnimallee 6, 14195 Berlin, Germany}

\author{Paolo A. Erdman \orcidlink{0000-0003-4626-2869}}
\affiliation{FU Berlin, Department of Mathematics and Computer Science, Arnimallee 6, 14195 Berlin, Germany}

\author{Zeno Sch{\"a}tzle \orcidlink{0000-0002-5345-6592}}
\affiliation{FU Berlin, Department of Mathematics and Computer Science, Arnimallee 6, 14195 Berlin, Germany}

\author{P. Bern{\'a}t Szab{\'o} \orcidlink{0000-0003-1824-8322}}
\affiliation{FU Berlin, Department of Mathematics and Computer Science, Arnimallee 6, 14195 Berlin, Germany}

\author{Frank Noé \orcidlink{0000-0003-4169-9324}}
\affiliation{FU Berlin, Department of Mathematics and Computer Science, Arnimallee 6, 14195 Berlin, Germany}
\affiliation{Microsoft Research AI4Science, Karl-Liebknecht Str. 32, 10178 Berlin, Germany}
\affiliation{FU Berlin, Department of Physics, Arnimallee 14, 14195 Berlin, Germany}
\affiliation{Rice University, Department of Chemistry, Houston, TX 77005, USA}

\date{20 June 2025} 
\begin{abstract}
    We propose a novel wave function partitioning method that integrates deep-learning variational Monte Carlo with ansätze based on generalized product functions. This approach effectively separates electronic wave functions (WFs) into multiple partial WFs representing, for example,  the core and valence domains or different electronic shells. Although our ansätze do not explicitly include correlations between individual electron groups, we show that they accurately reproduce the underlying physics and chemical properties, such as dissociation curve, dipole moment, reaction energy, ionization energy, or atomic sizes. We identify the optimal number of core electrons and define physical core sizes for Li to Mg atoms. Our results demonstrate that core electrons can be effectively decoupled from valence electrons. We show that the core part of the WF remains nearly constant across different molecules and their geometries, enabling the transfer and reuse of the core part in WFs of more complex systems. This work provides a general framework for WF decomposition, offering potential advantages in computing and studying larger systems, and possibly paving the way for ab-initio development of effective core potentials. Though currently limited to small molecules due to scaling, we highlight several directions for extending our method it to larger systems.
\end{abstract}


\maketitle

\section{Introduction} \label{sec:intro}

A central goal of quantum chemistry is to develop methods that accurately capture the correlation between all electrons in a molecule, while the nuclear motion is typically decoupled from the problem using the Born--Oppenheimer approximation. However, achieving exact solutions to the electronic Schrödinger equation is generally infeasible, necessitating the use of approximation methods, except for the simplest of systems. The most basic of these, the Hartree-Fock (HF) method, neglects direct electronic correlation by expressing the wave function as an antisymmetrized product of uncorrelated single-particle orbitals, effectively separating the all-electron wave function into single-particle functions (orbitals). While HF is computationally efficient, its accuracy is limited.  Nonetheless, it provides a valuable framework for understanding electronic orbitals, which play a crucial role in quantum chemistry and serve as the foundation for many more advanced methods that incorporate electron correlation. While post Hartree-Fock or multireference methods such as CCSD(T), CI, or CASSCF can yield highly accurate solutions, they are not generally applicable to all types of systems and there remains a cost-accuracy tradeoff, in that more accurate methods suffer from increasingly impractical scaling with the number of electrons. 
Moreover, most of these methods offer little insight into individual parts of a chemical system, only providing us with properties of the systems as a whole. 

Quantum chemical systems often exhibit regions with distinct characteristics -- for example, core regions with high density of high-energy electrons and valence regions where electrons move more slowly. Due to the indistinguishable nature of quantum particles, explicitly partitioning electrons into distinct groups and treating them separately poses a significant challenge. Nonetheless, separating valence and core electrons has been extensively studied. In simulations of systems with heavy elements, many quantum chemistry methods either leave the cores uncorrelated, fully frozen\cite{yu_accurate_2021}, or eliminated entirely using effective core potentials (ECP)\cite{burkatzki_energy-consistent_2007, bennett_new_2017}. 

In this work, we introduce a variational method that partitions the wave function into several independent groups of electrons, analogous to how the HF method divides an $N$-electron system into $N$ independent one-electron orbitals. Unlike the HF method, our approach preserves the correlations within each group, neglecting only inter-group correlations, thereby enabling us to treat each electron group separately. Our method inherently omits inter-group correlation energy, slightly increasing the total energy, when compared to an unrestricted simulation with a fully correlated ansatz. However, this inter-group correlation energy offset is approximately canceled out in ordinary chemical reactions where the electron groups are appropriately selected and nuclei are sufficiently separated. Other chemical properties, such as energy dissociation, reaction energy or ionization energy, are still accurately reproduced by our method.

We employ ansätze based on generalized product functions (GPF), first introduced by \citet{mcweeny_density_1959}, where the electronic wave function is formulated as an antisymmetrized product of multi-electron \textit{group functions}, rather than individual one-electron orbitals. They can thus better capture correlations within each electron group, unlike the HF method, offering deeper insights into electronic structure especially in systems where electrons can be meaningfully grouped, such as in atomic shells. There have been a number of studies employing GPF, for instance analyzing chemical bonds by partitioning electronic density and energy into interference and quasi-classical contributions\cite{cardozo_energy_2009, zhao_energy_2018}. However, almost all the works on GPFs have used a strong orthogonality condition that significantly restricts the variational space. We do not impose such a restriction, enabling the use of much more expressive ansätze.

We build our methods on deep QMC, i.e., variational Monte Carlo (VMC) in combination with deep-learning wave function ans\"atze, which have attracted significant interest in recent years due to advances in machine learning. Deep QMC is a relatively novel approach first introduced for bosonic lattices by \citet{carleo_solving_2017}. Deep QMC for fermionic wave functions of molecules has been pioneered by \citet{hermann_deep-neural-network_2020} and \citet{pfau_ab_2020} with many further improvements and extensions\cite{schatzle_deepqmc_2023, von_glehn_self-attention_2022, spencer_better_2020, li_fermionic_2022, scherbela_solving_2021, li_computational_2024, tang_deep_2025, schatzle_ab-initio_2025}, which makes deep QMC competitive with other state-of-the-art quantum chemistry methods. 
Opposed to traditional post Hartree-Fock methods, which construct expansions in Slater-Determinants of single particle orbitals, deep QMC wave functions explicitly introduce correlation to the electronic orbitals, making them a natural fit for grouping electrons.
We introduce the GPF approach to deep QMC, leveraging the expressive power of neural network wave functions to accurately represent different parts of a chemical system.

The paper is structured as follows. In section \ref{sec:methods}, we provide a brief description of deep QMC methods, followed by the introduction of a GPF-based ansatz that allows us to separate the all-electron problem into several distinct parts. In section \ref{sec:results}, we present our results, starting with different core-valence partitionings on small atoms from Li to Mg and determining the optimal number of core electrons along with the corresponding core and atomic radii. We then investigate the transferable properties of core wave functions across different molecules. Lastly, we present two examples of $\ce{Li_2}$ molecule and $\ce{Na}$ atom as cases where 3-way splitting is particularly well suited, and compute the ionization energy of $\ce{Na}$. Section \ref{sec:conclusion} summarizes our findings and concludes the work.

\section{Methods}\label{sec:methods}

\subsection{Deep Quantum Monte Carlo}

Deep QMC methods aim to accurately approximate the lowest eigenstate of an operator $\hat{H}$, which is typically an electronic Hamiltonian in the Born-Oppenheimer approximation describing an $N$-electronic system, but our approach is in principle not limited to such Hamiltonians. 
The VMC method is employed to variationally optimize the parametrized ansatz $\psi$ by minimizing its energy expectation value
\begin{equation}\label{eq:variational_optimization}
    E = \frac{\langle \psi | \hat{H}|\psi\rangle }{\langle \psi | \psi\rangle} = \mathop{\mathbb{E}}_{\mr\sim|\psi|^2}\left[ \frac{H\psi(\mr)}{\psi(\mr)} \right]\,, 
\end{equation}
where $\mr = (\br_1, \dots, \br_N)$ is an electronic configuration. In practice, the expectation value can be reliably estimated using Monte Carlo integration. 
Electronic configurations $\mr$ are generated from the many-body probability distribution $\propto \abs{\psi(\mr)}^2$ via Markov chains. Note that in VMC for molecules, electron spin can be treated implicitly by using spin-assigned wave functions. Details of the implementation can be found in \citet{schatzle_deepqmc_2023}.

For the parametrized ansatz $\psi$, we employ the Psiformer architecture\cite{von_glehn_self-attention_2022} based on self-attention, with several adjustments. Most notably, we use a scaled-down version of Psiformer as it only needs to describe simpler parts of the system rather than the full system. This also helps compensate for the fact that evaluating GPFs requires many inference passes; see below. We further adjust Psiformer's nuclear cusps and asymptotic behavior via nucleus-electron envelopes, and we restrict the input features to the neural network architecture when necessary. A detailed description of the ansatz and our modifications is in Appendix \ref{app:WF_ansatz}.

\subsection{Partitioned wave function ansatz}\label{ssec:splitWF}
We introduce a method to separate the full $N$-electron wave function $\psi(\mr)$ into several group wave functions $\psi_k$ each with $N_k$ electrons, so that $\sum_k^K N_k = N$. 
In general, it is not possible to split any wave function $\psi(\mr)$ into several independent electron groups, because the separation would remove explicit correlations between individual electron groups. However, by a careful selection of the particular electron groups we can achieve high accuracy, even if the explicitly-correlated WF is not reconstructed exactly.

We have temporarily set aside electron spin for clarity and we define the GPF-based wave function as
\begin{gather}
    \psi_\text{a}(\mr) = \mathcal{A}\left[ \psi_1 \dots \psi_K \right] (\mr) = \label{eq:wf_separation}\\
    = \sum_{\mathclap{\pi\in \text{intergroup}\hspace{-2ex}}} \text{sgn}(\pi) \psi_1(\br_{\pi(1)}, \dots, \br_{\pi(N_1)}) \dots \psi_K(\br_{\pi(N-N_K)}, \dots, \br_{\pi(N)})\,,\notag
\end{gather}
where $\mathcal{A}$ is a partial antisymmetrizer that only performs permutations $\pi$ between different groups (intergroup permutations). Exchanges of electrons within the same group are not needed, because each group-WF is antisymmetric, $\psi_k(\dots, \br_i, \br_j, \dots) = - \psi_k(\dots, \br_j, \br_i, \dots)$. Unlike previous works featuring GPFs\cite{mcweeny_density_1959, huzinaga_theory_1971, wilson_group_1976, colle_atoms_1984, cardozo_energy_2009, zhao_energy_2018}, we represent each group-WF $\psi_k$ with a neural network. We refer to the function \eqref{eq:wf_separation} as split-WF. As the VMC method does not require normalized wave functions, the split-WF is left unnormalized. 

The equation \eqref{eq:wf_separation} can be viewed as a natural generalization of a Slater determinant to many-body orbitals. When each group-WF contains exactly one electron, $K=N$, this formulation reduces to a standard Slater determinant. However, in other cases, it offers increased expressivity by explicitly accounting for electrons correlations within each group. We also note that in the case of $N_i = 2 ~\forall i$, taking a linear combination of split-WFs, with fixed $N_1, \dots, N_K$, would correspond to the theory of geminals\cite{surjan_introduction_1999}.




Unlike almost all previous works that employed GPFs \cite{mcweeny_density_1959, huzinaga_theory_1971, kahn__1976, szasz_pseudopotential_1968, colle_atoms_1984, cardozo_energy_2009, zhao_energy_2018}, we do not require the strong orthogonality condition that would impose a significant restriction on the group-WFs. As a consequence, our group-WFs are much more expressive. However, without the strong orthogonality condition, a decomposition of a split-WF $\psi_\text{a}(\mr)$ into group-WFs using formula \eqref{eq:wf_separation} is not unique. 
This aspect is discussed in the Appendix \ref{app:non_uniqueness}. This does not pose an issue for the wave function splitting itself, however it may pose an issue if we try to use the group-WFs to generate effective core potentials, see Appendix \ref{app:fitting_bfd}.


Lastly, we discuss how electron spin is handled.
Since the molecular Hamiltonian does not depend on spin, VMC studies typically work with spin-assigned real-valued wave functions $\psi(\mr^\uparrow, \mr^\downarrow)$, where $\mr^\uparrow = \br_1,\dots,\br_{N^\uparrow}$ are the coordinates of the first $N^\uparrow$ electrons that are treated as distinguishable from the remaining $N^\downarrow = N - N^\uparrow$ down-electrons denoted with $\mr^\downarrow = \br_{N^\uparrow+1},\dots,\br_{N^\uparrow+N^\downarrow}$. This trick fixes the $z$-component of the total spin $\sigma^z$. Consequently, the antisymmetrization of up and down electrons has to be done separately. The spinless formula \eqref{eq:wf_separation} is adapted to the following form to account for electron spin
\begin{align}
    \psi_\text{a}(\mr^\uparrow, \mr^\downarrow) = \mathcal{A}^\downarrow \left[ \mathcal{A}^\uparrow \left[  \psi_1 \dots \psi_K \right](\mr^\uparrow)\right](\mr^\downarrow)\,, \label{eq:wf_separation_spin}
\end{align}
where $\mathcal{A}^\uparrow$ is only applied to the up-electrons of each group wave function $\psi_k(\mr_k^\uparrow, \mr_k^\downarrow)$ and likewise for $\mathcal{A}^\downarrow$. The total number of terms in formula \eqref{eq:wf_separation_spin} is 
\begin{align}
    \frac{N^\uparrow !}{N_1^\uparrow ! \dots N_K^\uparrow !} \times \frac{N^\downarrow !}{N_1^\downarrow ! \dots N_K^\downarrow !}\,, \label{eq:antisymm_scaling}
\end{align}
where $N^\uparrow_i$ and $N^\downarrow_i$ denote the numbers of up and down electrons in $i$-th group respectively. It is necessary to evaluate all the group-WFs for each term, which poses a significant computational challenge when dealing with larger systems. Appendix \ref{sec:speedup} discusses potential solutions to this limitation. 

\subsection{Relation to effective core potential (ECP)}
The widely studied ECP methods\cite{bachelet_pseudopotentials_1982,burkatzki_energy-consistent_2007, trail_correlated_2015,trail_shape_2017,bennett_new_2017,annaberdiyev_new_2018} are related to the split-WF approach in that both provide wave functions for valence electrons which are not explicitly correlated to the core electrons.
The ECP approach approximates the effect of core electrons by introducing an additional ECP term included in the molecular Hamiltonian. 
This approximation enables the removal of the core electrons entirely from the simulation, thereby reducing computational cost through solving a simplified problem.
The construction of the ECP term typically relies on matching the electronic properties to reference data. 
In contrast, our partitioning approach introduced in Sec. \ref{ssec:splitWF} naturally provides access to a wave function of the valence electrons from first principles, being an \textit{ab initio} solution based on the unaltered all electron molecular Hamiltonian.

Although the wave functions yielded from ECP calculations have a fundamentally different origin than valence-WFs, they both describe valence electrons only. We therefore include a comparison of some of their properties in our investigations presented in the following section.
\section{Results} \label{sec:results}

Atoms and molecules are quantum systems that are naturally suited for separation into the core and valence groups of electrons. In this section, we first apply our WF splitting method with $K=2$ group wave functions to atomic systems, determining the optimal numbers of core and valence electrons, henceforth denoted as $N_\text{c}$ and $N_\text{v}$ respectively. We refer to the two group-WFs as a core-WF and a valence-WF for clarity. Using this decomposition, we compute the core and atomic radii. Next, we investigate the transferability properties of the core-WF on a simple system of the BeH molecule. Lastly, we present wave function splitting into $K=3$ group functions for the $\ce{Li2}$ molecule and the $\ce{Na}$ atom, and compute the sodium ionization energy.

All QMC calculations were performed using the \textsc{DeepQMC}\cite{deepqmc_2024} software package, which implements deep QMC along with many of its extensions.


\begin{figure}[tbh]
    \centering
    \includegraphics[width=0.95\linewidth]{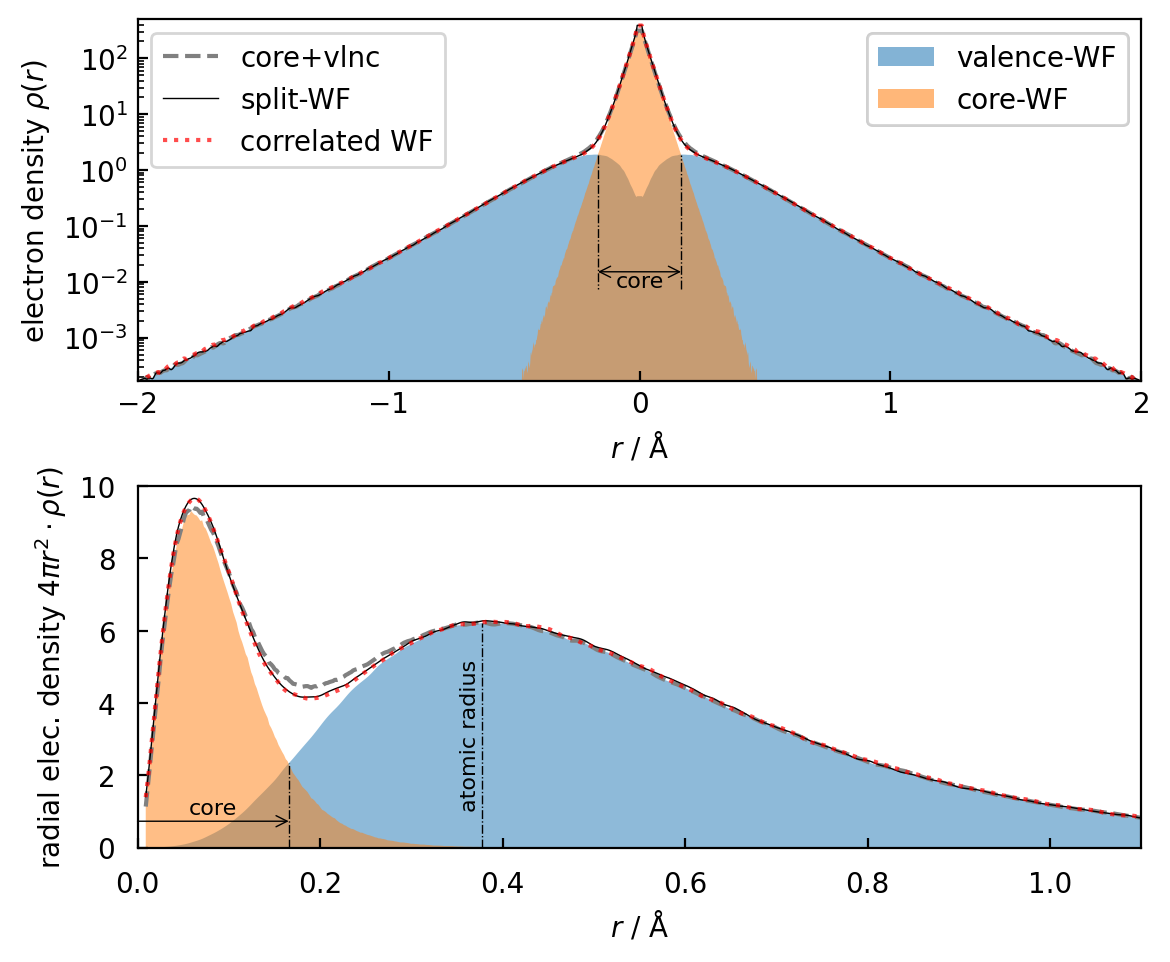}
    \caption{Example of one body electron density of the core-valence WF splitting for F atom with $\sigma^z = 1/2$. A setup with two core and seven valence electrons is used. Both absolute (top) and radial (bottom) density is shown. The core domain defined as region with higher density of core electrons than valence electrons is indicated. Atomic radius given by the maximum of the radial valence density is also indicated.}
    \label{fig:example_density}
\end{figure}

We begin the investigation with simple single-atom systems (see Appendix \ref{app:WF_ansatz} for experiment details).
Figure \ref{fig:example_density} shows a representative case of WF splitting for a fluorine atom. The one-body densities were calculated using the kernel density estimation method (detailed in Appendix \ref{app:shape}). The spatial distributions clearly demonstrate that both group-WFs occupy distinct domains with small overlap. The core-WF density accumulates near the nucleus, while the valence-WF density is suppressed in this region and extends into the wider valence domain. The split-WF density agrees very well with the reference obtained from a standard \textsc{DeepQMC} calculation. The sum of the valence-WF and core-WF densities closely matches the split-WF density, with a slight deviation in the core-valence transition region. This suggests a minor interference between both group-WFs, because the densities would be strictly additive only if there was zero overlap between individual group-WFs. 


\subsection{Optimal number of core electrons}

\begin{figure}[bth]
    \centering
    \includegraphics[width=0.99\linewidth]{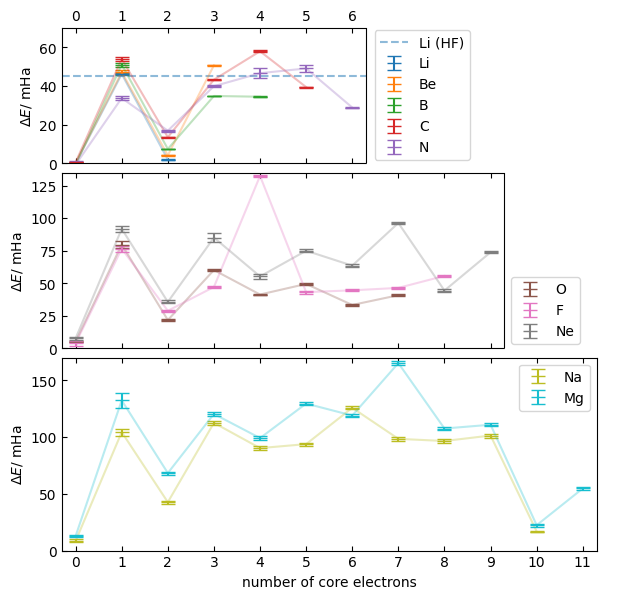}
    \caption{Energy error of core-valence splittings with various numbers of core electrons. The splitting is done such that the valence function has a spin of either 0 or $+{1}/{2}$ and the core has nonnegative integer spin so that their sum gives the ground state spin of the corresponding atom. Reference energies were taken from \citet{oneill_benchmark_2005} and HF energies were calculated with {\tt pyscf} package using complete basis set extrapolation limit. The dashed line shows the HF energy error of Li atom, while for heavier atoms, the HF energy errors exceed the plot range.}
    \label{fig:energy_errors}
\end{figure}

Electrons that are closest to the nucleus exhibit substantially different characteristics from the outermost valence electrons. However, in practice, choosing where to draw the line between electrons labeled as core and valence that best reflects the properties of the correlated wave function is not necessarily straightforward. The conventional approach is to analyze the one-electron orbital overlaps to identify which orbitals are more distinctly separated from the valence ones. In contrast, our method offers a way to independently determine the optimal number of core electrons without relying on orbital representation, making it potentially applicable to systems where orbitals are not well defined. The optimal number of core electrons is naturally identified as the one that minimizes cross-correlation error -- that is, it achieves the lowest variational energy. Figure \ref{fig:energy_errors} shows the errors of variational energy for different core sizes $N_\text{c}$ in light atoms ranging from Li to Mg.

For the first row atoms, the results show that the optimal number of core electrons is 2. Although cores with 2 electrons yields good variational energies for Na and Mg as well, the best number is 10 for those two atoms. This is consistent with the core sizes commonly used in ECP methods\cite{burkatzki_energy-consistent_2007,bennett_new_2017}. As expected, the split-WF energies are always lower than the HF energies for all atoms and all core sizes because HF does not account for electronic correlations directly. Likewise, the regular explicitly-correlated WF denoted as $N_\text{c}=0$ always reaches significantly higher accuracy in total energies. We later demonstrate that the split-WF ansatz still accurately describes the chemically relevant properties of the quantum system, despite not achieving the lowest variational energy. 

If the same architectural configuration were used for both core and valence WF ansätze, Figure \ref{fig:energy_errors} would exhibit symmetry such that a split-WF with $N_\text{c}$ core electrons would yield the same variational energy as one with $N-N_\text{c}$ core electrons\footnote{Except for C, N and O atoms that have at least two unpaired electrons. For such atoms, $N'_\text{c} = N - N_\text{c}$ would result in different spins of the core and valence WFs.}. In such cases, we observed that it is essentially random which of the two group functions converges to represent the core versus valence region. We decided to use slightly different architecture for both group-WFs, details of which are in Appendix \ref{app:WF_ansatz}. We found that the most important factor determining which group-WF becomes core and which one becomes valence is the initialization of the envelope width (we use narrower envelope initialization for core- and wider for valence-WF). The width initialization also significantly affects the stability and convergence speed, so it is important to start with reasonable guesses. This also applies to conventional deep-QMC calculations. Although the envelope width can be learned during optimization, we have never observed a swap of core- and valence-WF domains once there is a hint of core-valence separation during training, even when such a swap could lower the energy. This explains why cases with 2 valence electrons did not achieve the same level of accuracy as those with $2$ core electrons.

It is interesting to observe that for atoms with an even number of electrons (most notably O and Ne), the split-WF systematically reaches lower energies when the number of core electrons is even, resulting in a zigzag pattern. This finding indicates that core-WF with zero spin is advantageous, which is in agreement with the orbital picture of paired electrons, because suppressing the correlation within an orbital electron pair leads to a larger error. Furthermore, the results recover the theory of electron shells from first principles, with full shells providing the best separation. We also note that it is difficult for split-WF ansatz learn the spherical symmetry of the single-atom systems, which is further detailed in Appendix \ref{app:spherical_symm}.

\subsection{Physical core and atomic radii}
\begin{figure}[thb]
    \centering
    \includegraphics[width=0.72\linewidth]{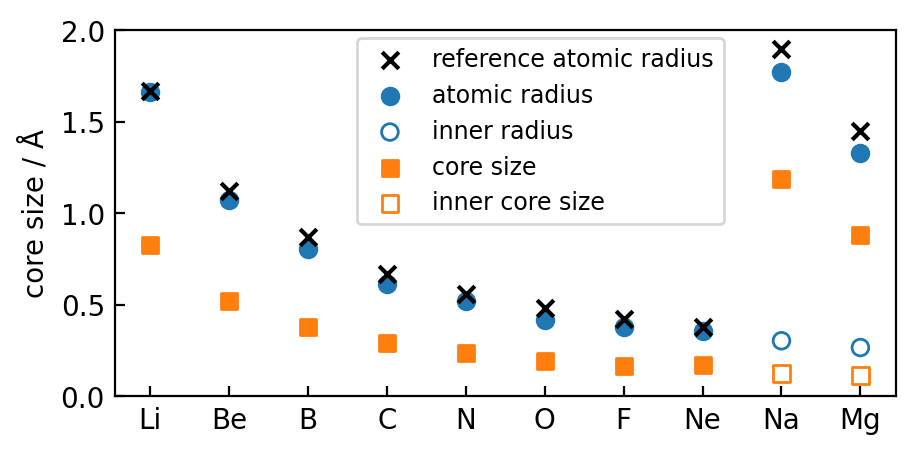}
    \caption{Core sizes and atomic radii for Li to Mg atoms. Reference atomic radii are taken from \citet{clementi_atomic_1967}. An inner core size and inner radius is also shown for Na and Mg, which was calculated using 2 core electrons instead of 10.}
    \label{fig:core_sizes}
\end{figure}

Although the concept of a core region is physically intuitive, its exact spatial extent is not straightforward to define. However, once the atomic wave function is separated into the core and valence groups, we can define the border between these two regions as the area where both the core-WF and the valence-WF have the same electronic density (see Figure \ref{fig:example_density} for an illustration). In this way, the atomic core radii can be estimated, see Figure \ref{fig:core_sizes}. We also compute the atomic radii following the definition of \citet{clementi_atomic_1967} as the radii of maximum electron density in the outermost atomic shell. In our approach, this corresponds to the maximum of valence-WF density when $N_\text{c}=2$ for first-row atoms and $N_\text{c}=10$ for Na and Mg.

Both the atomic and core radii decrease across each period. This is due to increase in the positive nuclear charge attracting the valence electrons while the number of core electrons screening the nucleus remains constant. The atomic radii obtained in our calculations are systematically smaller than those reported in the reference, because our method more accurately accounts for electron correlation compared to the SCF method with a minimal basis set used in the reference. In order to move closer to the nucleus and reduce the total energy, valence electrons must account for the repulsion from other electrons, which requires an accurately correlated description of their motion.

\subsection{Transferability of the core-WF across molecules} \label{ssec:transferrable_properties}

The formation of molecules from atoms is primarily driven by the interactions among valence electrons.
If valence electrons and core electrons are well separated, the description of chemical bonding may involve changes to the valence wave function only. 
To probe this assumption, we study the transferability of the core-WF across different molecular geometries during bond dissociation.
We first examine errors due to decoupling the core and valence parts of the wave function, training independent split-WF ansatzes at each geometry.
Subsequently, we investigate the additional approximation arising from reusing the core-WF trained on a single atom in the process of chemical bonding.

\begin{figure}[tbh]
    \centering
    \includegraphics[width=0.8\linewidth]{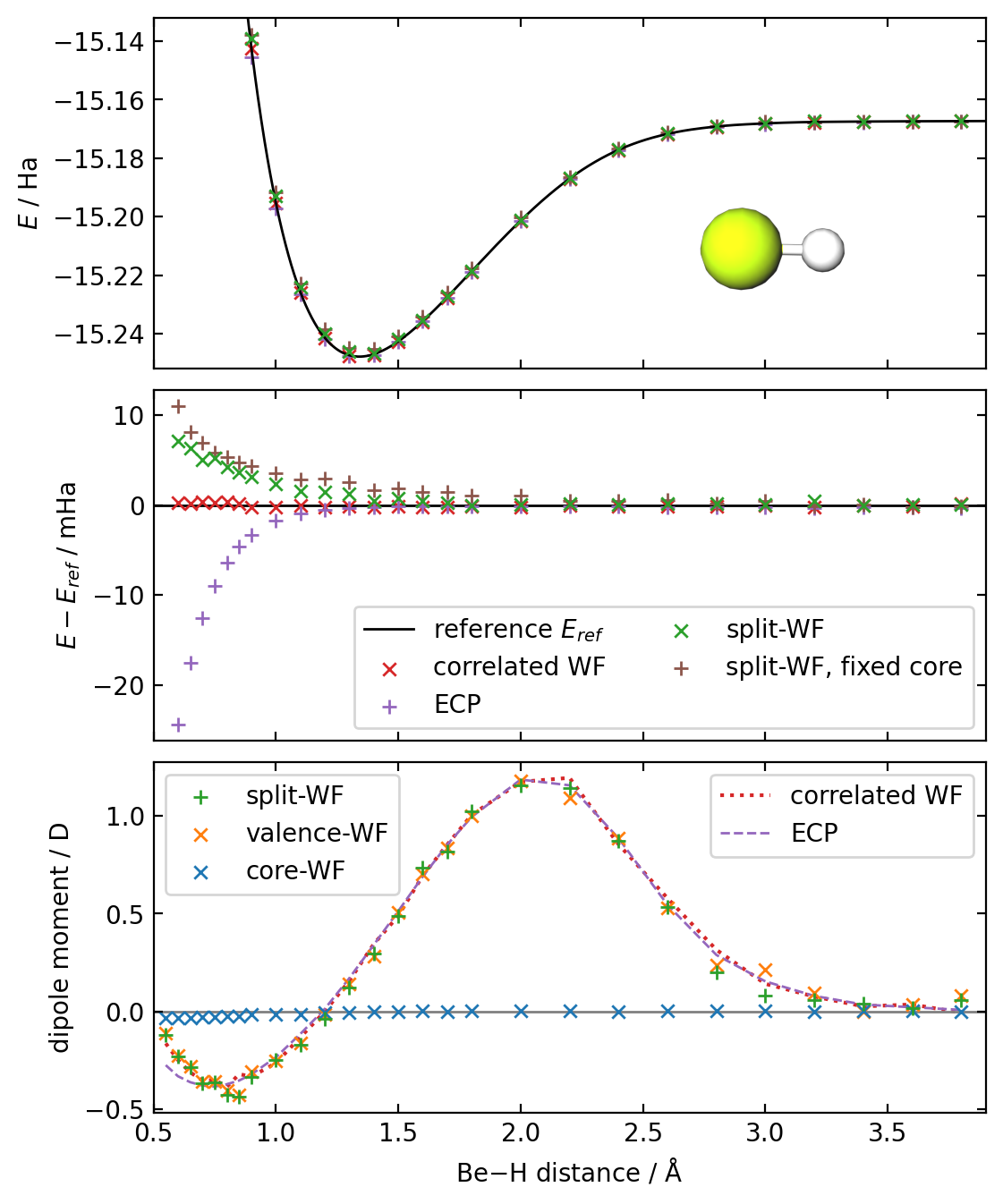}
    \caption{Dissociation curve and dipole moment of BeH comparing the results for conventional correlated WF (red), split-WF ansatz (green), and split-WF ansatz with fixed core-WF taken from calculation on a single Be atom (brown). The energies of both cases of split-WF are shifted by a constant factor based on the stretched geometry. Reference BeH energies were taken from \citet{koput_ab_2011}. Energies of \textsc{DeepQMC} calculations employing effective core potential\cite{bennett_new_2017} for Be nucleus are also shown (purple) and are shifted by a constant factor determined form an ECP calculation of single Be atom. Dipole moments were also computed for valence (orange) and core (blue) parts of the trained split-WFansatz.}
    \label{fig:BeH_fixedBeCore}
\end{figure}

Since the split-WF ansatz neglects core-valence correlation, its variational energies are slightly offset with respect to the fully correlated ansatz conventionally used in {\sc DeepQMC}.
Here, we empirically show that the energy offset introduced by employing the split-WF ansatz is almost constant across different molecular geometries, and cancels out when computing relative energies. We chose 25 different bond lengths of the BeH molecule. For each geometry, we performed an independent optimization of the split-WF. The results are shown in figure \ref{fig:BeH_fixedBeCore}. The offset in total energies is about $4\,\text{mHa}$ and is constant for stretched geometries. The energy error increases when both atoms are very close together, indicating that the squeezed geometries have an increased correlation between the core and the valence electrons that the split-WF ansatz cannot account for. 

Figure \ref{fig:BeH_fixedBeCore} also shows that the dipole moment obtained with the split-WF closely follows the correlated calculation throughout the dissociation, with accuracy comparable to that of the ECP calculation. Additionally, the dipole moment can be computed for individual group functions using the corresponding core/valence electron density. The valence dipole moment aligns with the all-electron results, and the core dipole moment is zero with a slight systematic shift at compressed geometries showing that the core electrons can respond to the proximity of valence electrons.

\begin{figure}[tbh]
    \centering
    \includegraphics[width=1.0\linewidth]{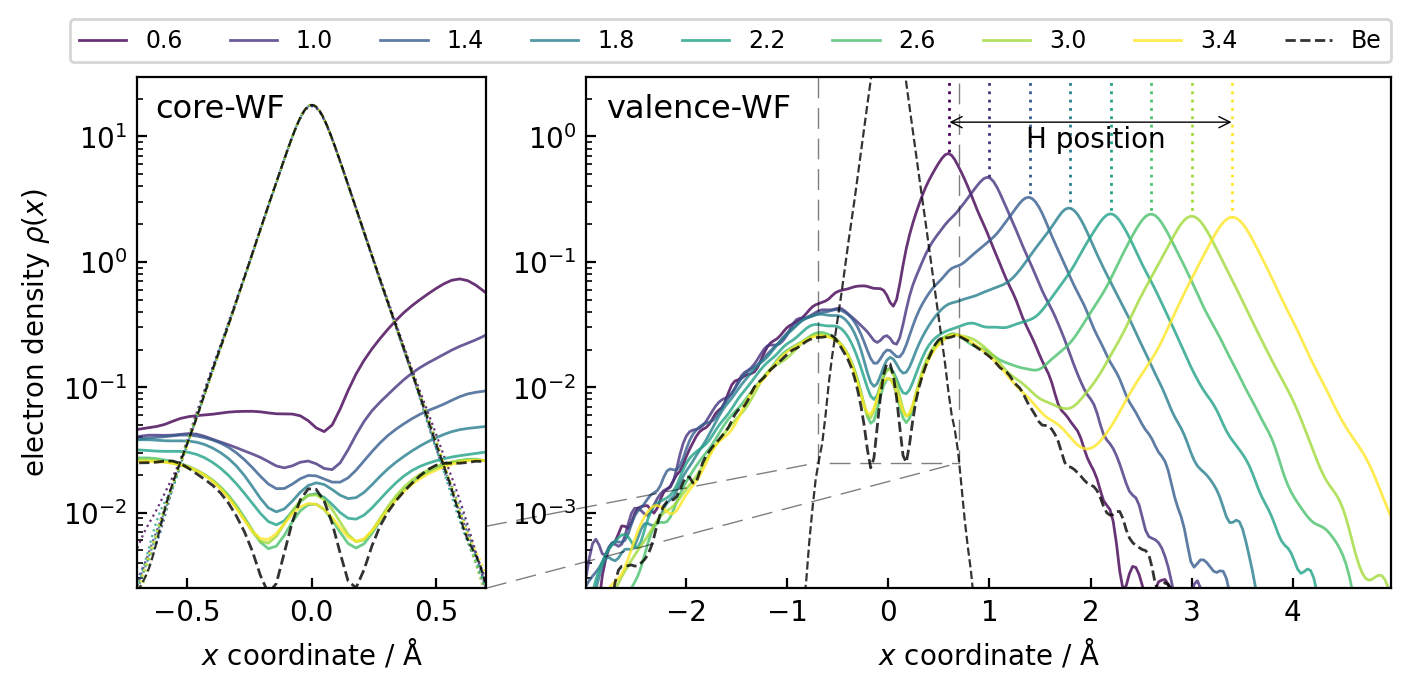}
    \caption{One body density of core and valence wave functions of seven independent {\sc DeepQMC} calculations with different BeH bond lengths. Valence densities are shown on both plots while the core density is only shown in the zoomed in left panel. Coordinates of H atom are denoted with dashed lines and the bond distance is given in legend in Angstroms. For comparison, densities from calculation of single beryllium atom are also shown with black dashed lines in both plots. In the left plot, all the core densities are overlapping, forming a single line. In contrast, the valence-WF density is changing as the hydrogen core moves.}
    \label{fig:BeH_different_geometries}
\end{figure}

To visually confirm that the core-WF is constant for different molecular geometries, we plot its density for different bond lengths of the BeH molecule in Figure \ref{fig:BeH_different_geometries}. The core-WF densities for all geometries closely overlap, and the changes in hydrogen coordinate are only reflected in the shape of the valence-WF where the maximum of the density tracks the location of the H atom.

To take this step further and validate that the core-WF indeed remains constant, we took the core-WF from the split-WF calculation on a single Be atom and applied it to the same set of geometries of the BeH molecule. We fixed the parameters of the \textit{pretrained} core-WF and only optimized the valence-WF parameters. An independent valence-WF optimization was performed for each geometry and the results are shown in Figure \ref{fig:BeH_fixedBeCore}. The energies are in agreement with the energies from the previous experiment where both group-WFs were optimized with only a slightly increased errors for squeezed geometries. This indicates that core-WF can be effectively transferred between different molecules (and different geometries) without compromising accuracy of the resulting split-WF. This result also supports the general concept that molecular properties are mainly determined by the valence electrons.

\subsection{Reaction energy}

To further show that the error introduced by not explicitly accounting for the correlations between the core and valence electrons is not significant and that split-WF reproduces correct physics, a reaction energy of $\ce{CH2 + H2 \rightarrow  CH4}$ was computed. The split-WF ansatz reproduced the reaction energy with an error of $1.6 \pm 0.3\,\text{mHa}$ with respect to the reference energies taken from \citet{oneill_benchmark_2005}. The energy was computed as a weighted average of five independent calculations with different random seeds. When the core-WF was taken from a single C atom calculation and only the valence-WF was optimized, the error increased to $3.2 \pm 0.3\,\text{mHa}$. For comparison, the conventional explicitly correlated ansatz using the same adjusted Psiformer architecture yielded closer agreement with an energy error of $0.6 \pm 0.1\,\text{mHa}$.


\subsection{3-part splitting of \texorpdfstring{$\text{Li}_\text{2}$}{Li2} molecule}

\begin{figure}[bt]
    \centering
    \includegraphics[width=0.8\linewidth]{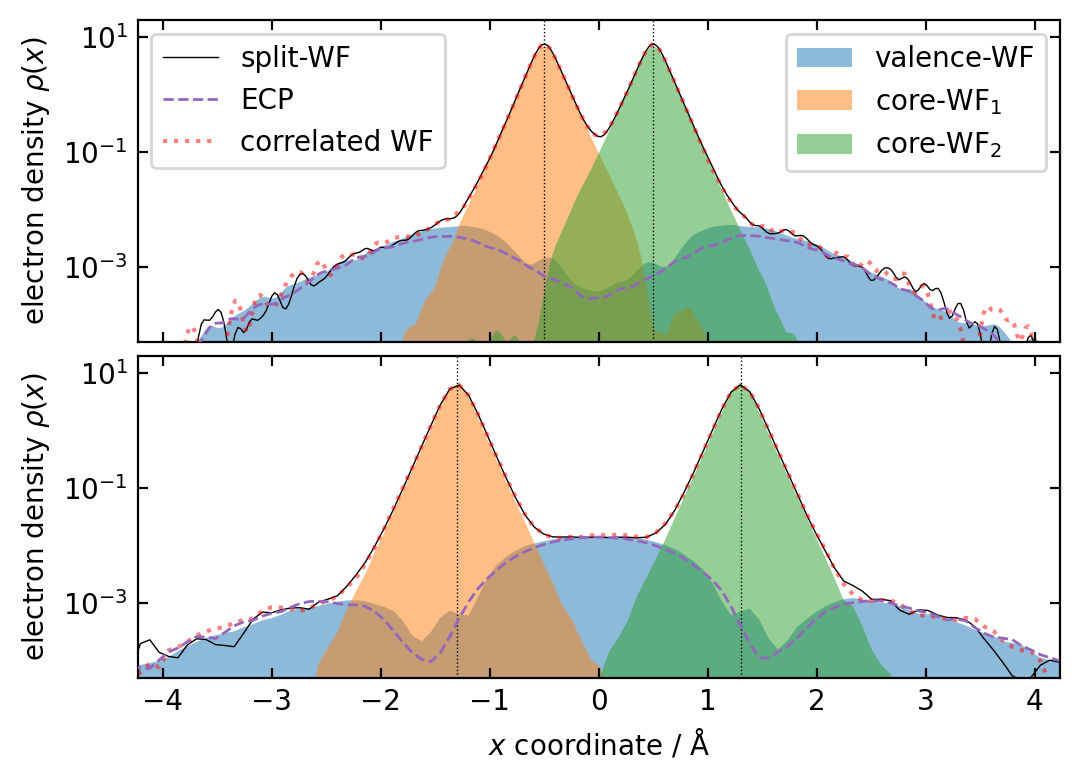}
    \caption{One body density of two different bond lengths of $\ce{Li_2}$ molecule with 2 core electrons at each lithium core and 2 valence electrons. Density of split-WF is compared with reference density and density of the wave function with both cores replaced by ECPs.}
    \label{fig:Li2_c2c2v2}
\end{figure}

Here we decompose the $\ce{Li2}$ molecule to two distinct core-WFs and one valence-WF. It is important to note that both core-WFs have modified input features such that they can explicitly \textit{see} only their own nucleus and not the other one. Figure \ref{fig:Li2_c2c2v2} shows that the split-WF density matches the reference density and that both cores are clearly separated from the valence-WF that describes the wave function at greater distances as well as in the space between the atoms.

\begin{figure}
    \centering
    \includegraphics[width=0.8\linewidth]{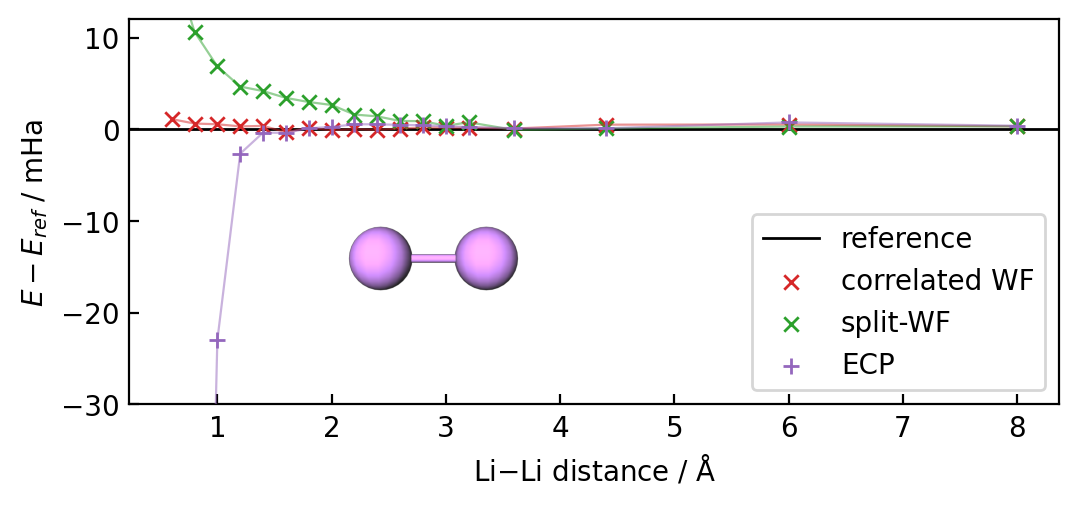}
    \caption{Energy error for $\ce{Li2}$ molecule. We compare the results of split-WF ansatz (green) with the explicitly correlated ansatz (red) and a wave function with effective core potential (purple), computed within \textsc{DeepQMC} framework. Due to a lack of good quality reference data from literature, an accurate \textsc{DeepQMC} calculation with larger ansatz, increased batch size and longer training was taken as a reference (black line).}
    \label{fig:Li2_dissoc}
\end{figure}

We plot the dissociation energy error of the $\ce{Li2}$ molecule in Figure \ref{fig:Li2_dissoc} to show that the physics is correctly reproduced when both Li cores are treated separately. As in the case of BeH, the split-WF produces a significant energy error only in very squeezed geometries. ECP energy is more accurate at bond distances greater than $1.4\,\textup{\AA}$, however, the ECP approach breaks down for very squeezed geometries, producing huge energy errors when lithium atoms are closer than $1\,\text{\AA}$.

\subsection{3-part splitting of Na atom}

\begin{figure}[htb]
    \centering
    \includegraphics[width=0.8\linewidth]{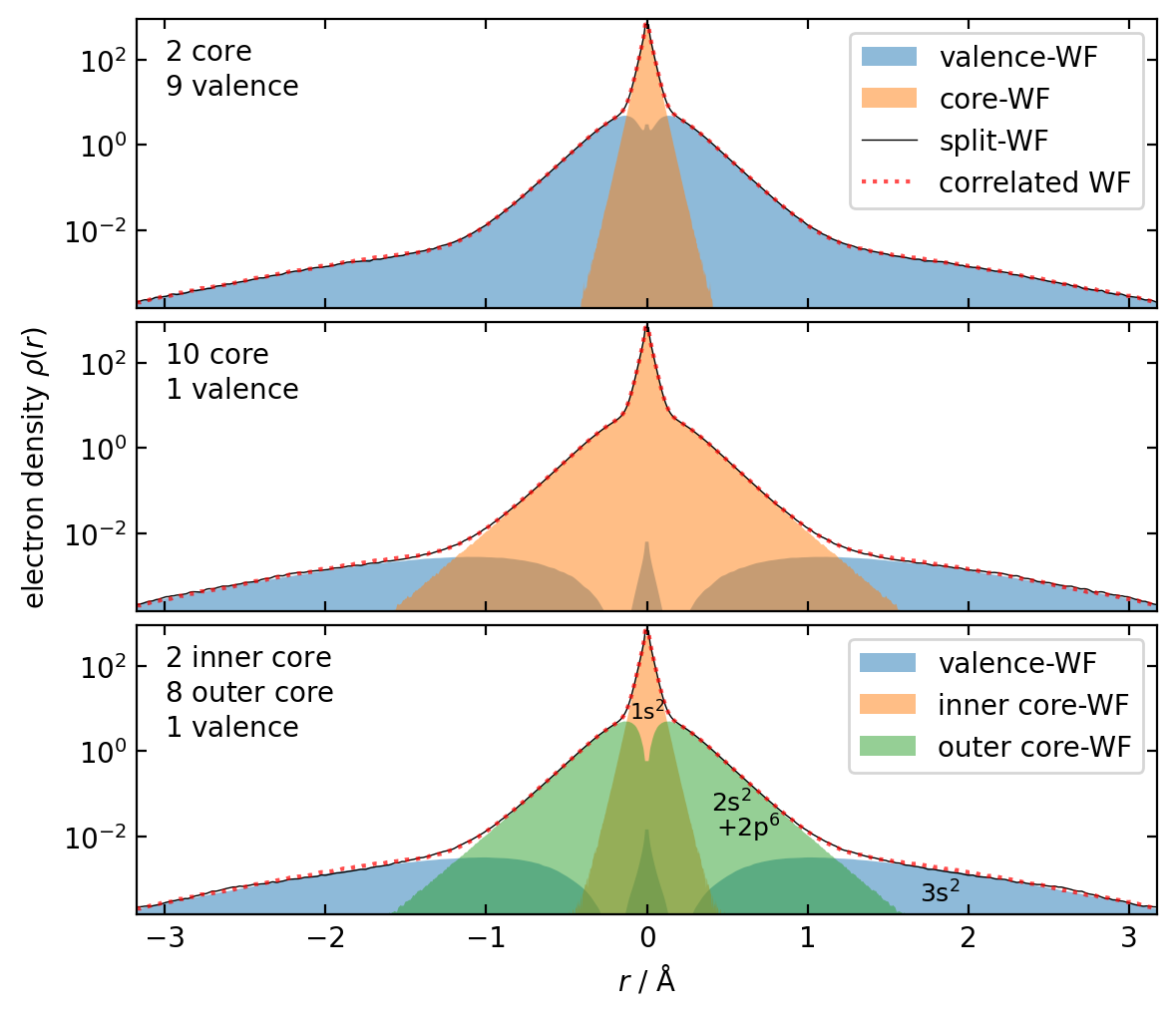}
    \caption{One body electron density of Na atom ground state. Three different ways of splitting the wave function are depicted. 2-part splitting with 2 core electrons is shown in the upper plot and with 10 core electrons is in the middle plot. The 3-part splitting (lower plot) effectively separates the system into individual electron shells of 2, 8 and 1 electrons. Numbers of up and down electrons for each decomposition are shown on the left.}
    \label{fig:Mg_c10v2}
\end{figure}

The sodium atom is a suitable system for 3-part splitting, as it contains 3 different electron shells -- inner core ($1\text{s}^2$) with two electrons, outer core ($2\text{s}^22\text{p}^6$) with 8 electrons and valence region ($3\text{s}^1$) with one electron. The result of the splitting is shown in the bottom panel of Figure \ref{fig:Mg_c10v2}. It shows that each group-WF naturally corresponds to a specific electron shell with a small overlap between different group-WFs.

We note that the valence-WF density has an unexpected peak at the origin. However, since the density of the core-WF is many orders of magnitude higher there, the value of the split-WF at $r=0$ is dominated by the contribution of the inner core-WF. In other words, the valence-WF value around the origin is irrelevant as long as it is much lower than core-WF. The fact that valence-WF values around the origin are arbitrary is not an issue for our work, but may become important if one would like to generate ECP from valence-WF, see Appendix \ref{app:fitting_bfd}.

To show that the physics is still correctly reproduced by the split-WF, we compute ionization energy of sodium, see table \ref{tab:ionization_energy}. Both the 3-part and 2-part split-WF ansätze show a fair agreement with the accurate \textsc{DeepQMC} calculation treated as a reference. Moreover, they agree slightly better than the explicitly correlated ansatz (correlated WF) that used the same number of training steps and samples. This shows that ionization has little effect on both the K and L shell and there is insignificant correlation between an M-shell electron and the core electrons.


\begin{table}[htb]
    \centering
    \begin{tabular}{ll}
    \toprule
        ansatz \hspace{1cm} & IE / mHa \\\midrule
        split-WF$_2$        & $185.3(11)$ \\
        split-WF$_3$        & {$184.8(12)$} \\
        correlated WF       & $191.5(31)$ \\
        baseline WF         & $186.5(14)$ \\
        experiment          & $188.8$ \\
    \bottomrule
    \end{tabular}
    \caption{Na ionization energy (IE). The split-WF ansatz was used with 2 core and 8 valence electrons for $\ce{Na+}$ calculation and either 2+9 electrons (split-WF$_2$) or 2+8+1 electrons (split-WF$_3$) for the computation of the neutral $\ce{Na}$ atom containing three electron shells. For comparison, we show IE computed using adjusted Psiformer (correlated WF) and conventional Psiformer with increased batch size and training time (baseline WF). The split-WF energy was computed as a weighted average of five independent calculations with different random seeds with occasional need to remove outliers to achieve stable result.}
    \label{tab:ionization_energy}
\end{table}
\section{Conclusion} \label{sec:conclusion}

A novel ab-initio method for partitioning many-electron wave functions was developed by combining the GPF-based ansatz with deep-learning wave functions. This method was applied to a number of small atoms and molecules successfully separating them into a narrow inert core part and a chemically relevant valence part. 
We were able to empirically discover chemical concepts about the electronic structure from first principles and show that molecular properties are mostly governed by the valence part of the split-WF. 
We confirmed that the optimal WF splitting consists of 2 core electrons for first-row atoms and 10 core electrons for Na and Mg, which is in line with the conventional orbital picture. These numbers also match the commonly used numbers of core electrons in ECP literature \cite{burkatzki_energy-consistent_2007,bennett_new_2017}. 
The core-valence splitting allowed us to define a boundary between core and valence domains. We reported the core and atomic radii of atoms from Li to Mg.

Although the split-WF ansatz is inherently less expressive than conventional ansätze that directly incorporate correlations between core and valence electrons, we showed that the resulting energy offset of split-WF remains constant except for highly compressed geometries, where correlations between core electrons and electrons of another atom become more significant. Consequently, the relative energies and other chemical properties remain accurate for most systems.

We further confirmed that the core part of the wave function is approximately constant for a given element across different molecules and their geometries, and we showed that it can be transferred between different systems with only a minor loss of accuracy. This allows us to simplify the all-electron variational problem into learning the valence part only, without having to optimize the core part. However, to this date the unfavorable scaling of the antisymmetrizer makes a general application of this technique impractical. Possible ways to mitigate combinatorial scaling are discussed in the appendix \ref{sec:speedup}. Alternatively, it is conceivable that antisymmetrization could be avoided by fitting a new ECP based on the core and/or valence wave function obtained with the WF partitioning method. An example of fitting an ECP based on an existing valence-type wave function is presented in the appendix \ref{app:fitting_bfd}. 

Lastly, we show that our method can split the electronic wave function into more than two parts. We presented an example of a $\ce{Li_2}$ molecule, where two Li cores are identified and treated separately. An example of a sodium atom was shown, where the wave function was successfully decomposed into three individual electronic shells, preserving the electronic correlation within each shell well. Computations of $\ce{Li2}$ dissociation and $\ce{Na}$ ionization energy confirmed that three-way splitting reproduces the correct physical properties with good accuracy.

In summary, combining the GPF-based ansatz with deep-learning wave functions gives us a novel tool to analyze the electronic structure of atoms and small molecules, and enables the removal of correlation between groups of electrons, generating suitable approximations for down stream tasks such as transferable wave function components and ECP fitting.

\begin{acknowledgments}
Funded by the Deutsche Forschungsgemeinschaft (DFG, German Research Foundation) under Germany's Excellence Strategy – The Berlin Mathematics Research Center MATH+ (EXC-2046/1, project ID: 390685689) projects AA2-20, AA1-6 and AA2-8.
The authors gratefully acknowledge the computing time made available to them on the high-performance computer \textit{Lise} at the NHR center NHR@ZIB. This center is jointly supported by the Federal Ministry of Education and Research and the state governments participating in the NHR (\url{www.nhr-verein.de}). The calculations carried out in this work were conducted using compute resources from the bec00268 project. MM used AI-assisted technology to improve writing style and check grammar and spelling. The authors carefully reviewed and edited the content as needed and assume full responsibility for the content of the publication.

\end{acknowledgments}

\section*{Data availability}
The data that support the findings of this study are available from the corresponding author upon reasonable request.

\bibliography{references}

\appendix
\section{Antisymmetrizer}\label{app:non_uniqueness}
In this appendix, we analyze key properties of our antisymmetrization scheme. In section \ref{sec:non_uniquness_problem}, we show that the group-WFs are not unique, section \ref{sec:speedup} discusses possible strategies to mitigate the combinatorial scaling, and in section \ref{sec:without_anti} we propose an alternative WF partitioning.

\subsection{Non-uniqueness of the WF partitioning} \label{sec:non_uniquness_problem}
Here, we present a proof that using the antisymmetrizer $\mathcal{A}$ to split the wave function into core and valence part does not produce unique group-WFs. Specifically, we show that one can alter a group-WF without affecting the resulting split-WF in any way.

Consider an antisymmetric combination of two antisymmetric spinless functions $f_{1\dots N_f}: \mathbb{R}^{N_f} \rightarrow \mathbb{R}$ and $g_{1\dots N_g}: \mathbb{R}^{N_g} \rightarrow \mathbb{R}$
\begin{equation}
    \psi_{1\dots N} = \mathcal{A}\left[f g\right]_{1\dots N}\,,
\end{equation}
where $N_f+N_g = N$ and we have introduced a shorthand notation for function arguments represented by lower indices $f_{1\dots N_f} := f(r_1, \dots, r_{N_f})$. The group functions $f$ and $g$ may be understood as core and valence. This proof assumes that $N_g \geq N_f$ and that $N_f$ is odd. The other cases are discussed later.

Let us define a third antisymmetric function $\Tilde{f}: \mathbb{R}^{N_g} \rightarrow \mathbb{R}$ acting on the same domain as $g$. We now modify the group function $g$ by adding this function, $g'_{1\dots N_g} = g_{1\dots N_g} + \Tilde{f}_{1\dots N_g}$. The modified total split-WF becomes
\begin{align}
    \psi'_{1\dots N} &= \mathcal{A} [f g']_{1\dots N} = \mathcal{A} [f \left( g + \Tilde{f} \right)]_{1\dots N} \\
    &= \underbrace{\mathcal{A} [f g]_{1\dots N}}_{\psi_{1\dots N}} + \mathcal{A}[f\Tilde{f}]_{1\dots N}\,. 
\end{align}
Thus, the split-WF remains unchanged after modifying $g \rightarrow g'$ iff
\begin{align}\label{eq:zero_antisymmetrizer}
    \mathcal{A}[f \Tilde{f}]_{1\dots N} = 0\,,
\end{align}
and this condition is fulfilled by function
\begin{equation}
    \Tilde{f}_{1\dots N_g}=\mathcal{A}[f h]_{1\dots N_g}\,,\label{eq:s1_extended_core_function}
\end{equation}
where $h_{1\dots(N_g-N_f)}$ is an arbitrary antisymmetric function. We show that \eqref{eq:s1_extended_core_function} satisfies \eqref{eq:zero_antisymmetrizer} in the following two paragraphs.


The condition \eqref{eq:zero_antisymmetrizer} contains two nested antisymmetrizers, $\mathcal{A}[f \mathcal{A}[f h]]_{1\dots N}$, which can be written as a sum of products of type $f_\pi f_\sigma h_\tau$ corresponding to different ways of distributing the $N$ indices (i.e. arguments) among the $f$, $f$ and $h$ functions. We denote $\pi, \sigma$ and $\tau$ three disjoint ordered tuples of indices representing partial permutations of $N_f, N_f$ and $N_g$, respectively. We use uppercase letters $\Pi, \Sigma$ and $\Tau$ to denote (unordered) sets of $N_f, N_f$ and $N_g$ indices, respectively. Since the antisymmetrizers contain all the inter-group permutations and no intra-group permutations, the expression \eqref{eq:zero_antisymmetrizer} is a sum of all unique divisions of $N$ indices into disjoint sets $\Pi, \Sigma$ and $\Tau$. 
We can see that for each division into $\Pi, \Sigma$ and $\Tau$, there is exactly one other division into $\Pi', \Sigma'$ and $\Tau$, such that $\Pi' = \Sigma$, $\Sigma' = \Pi$ and $\Tau$ remains unchanged. Since the functions $f$ and $h$ are antisymmetric, the ordering within each set only affects the sign, and the sum in \eqref{eq:zero_antisymmetrizer} is thus composed of pairs of terms $f_\pi f_\sigma h_\tau$ and $f_{\pi'} f_{\sigma'} h_\tau$ with the same absolute value.

Using the definition \eqref{eq:wf_separation}, we can determine the sign of the term $f_\pi f_\sigma h_\tau$ as $\text{sgn}(\pi\sigma\tau)$, where the juxtaposition of tuples denotes concatenation. Thanks to the assumption that $N_f$ is odd, we can write $\text{sgn}(\pi\sigma\tau) = -\text{sgn}(\sigma\pi\tau) = -\text{sgn}(\pi'\sigma'\tau)$. The terms within each pair therefore have opposing signs and the whole sum adds up to zero, proving the equation \eqref{eq:zero_antisymmetrizer}. 

Eqs. \eqref{eq:zero_antisymmetrizer} and \eqref{eq:s1_extended_core_function} can be interpreted as a generalization of the fact that a Slater determinant of one-particle functions is zero when two functions are linearly dependent, except we employ $N_f$-particle functions instead of one-particle orbitals. The case when $N_f$ is even cannot be handled by the same argument as both the above-mentioned terms would have the same sign (cf. swapping of an even number of rows in a Slater determinant results in no sign change). Nevertheless, for some specific forms of $f$, such as $f_{12} = h_1-h_2$, it is still possible to find $\Tilde{f}$ satisfying the condition \eqref{eq:zero_antisymmetrizer}. 

However, the case where $N_f$ is even is less relevant for our application. We mostly work with core-WFs of 2 electrons that possess 1 up and 1 electron. Therefore, conditions $N_\text{v}^s \geq N_\text{c}^s$ and $N_\text{c}^s$ is even usually hold for both spin parts, $s \in \{\uparrow,\downarrow\}$.

\subsection{Overcoming the combinatorial scaling of antisymmetrizer evaluation} \label{sec:speedup}
From the electron density plots (Figs. \ref{fig:example_density}, \ref{fig:Li2_c2c2v2}) we observe that the individual group-WFs typically describe spatially separated regions, most often a compact core region and diffuse valence region. The electron density of core-WF is localized near the nucleus and decays rapidly with distance. As a result, most of the terms in \eqref{eq:wf_separation_spin} are very small and the split-WF value is dominated by a small subset of terms terms in which near-nucleus electrons are assigned to the core-WF and distant electrons to the vlnc-WF. We verified empirically that, in trained split-WFs, the magnitudes of individual terms in \eqref{eq:wf_separation_spin} vary greatly in magnitude. Importantly, their relative magnitudes can be estimated directly from the electron-nucleus distances, without needing to evaluate the group-WFs themselves. This suggests a practical strategy to mitigate the prohibitive combinatorial scaling in eq. \eqref{eq:antisymm_scaling}. We could evaluate only the dominant terms in eq. \eqref{eq:wf_separation_spin}, and neglect the remaining terms whose values are expected to be close to zero based on the electron-nucleus distance. 

A simple approximation involves labeling the electrons as core and valence based on their distance from the nucleus. A first order approximation allowing only one core-valence swap would lead to a simple quadratic scaling 
$$\frac{N^\uparrow!}{N_\text{c}^\uparrow!N_\text{v}^\uparrow!}\times \frac{N^\downarrow!}{N_\text{c}^\downarrow!N_\text{v}^\downarrow!} \rightarrow N_\text{c}^\uparrow N_\text{v}^\uparrow + N_\text{c}^\downarrow N_\text{v}^\downarrow\,$$
greatly reducing the number of necessary group-WF evaluations. We find that second-order terms (those including two swaps) have typically 6 orders of magnitude lower absolute value than first-order terms.

\begin{figure}[htb]
    \centering
    \includegraphics[width=0.90\linewidth]{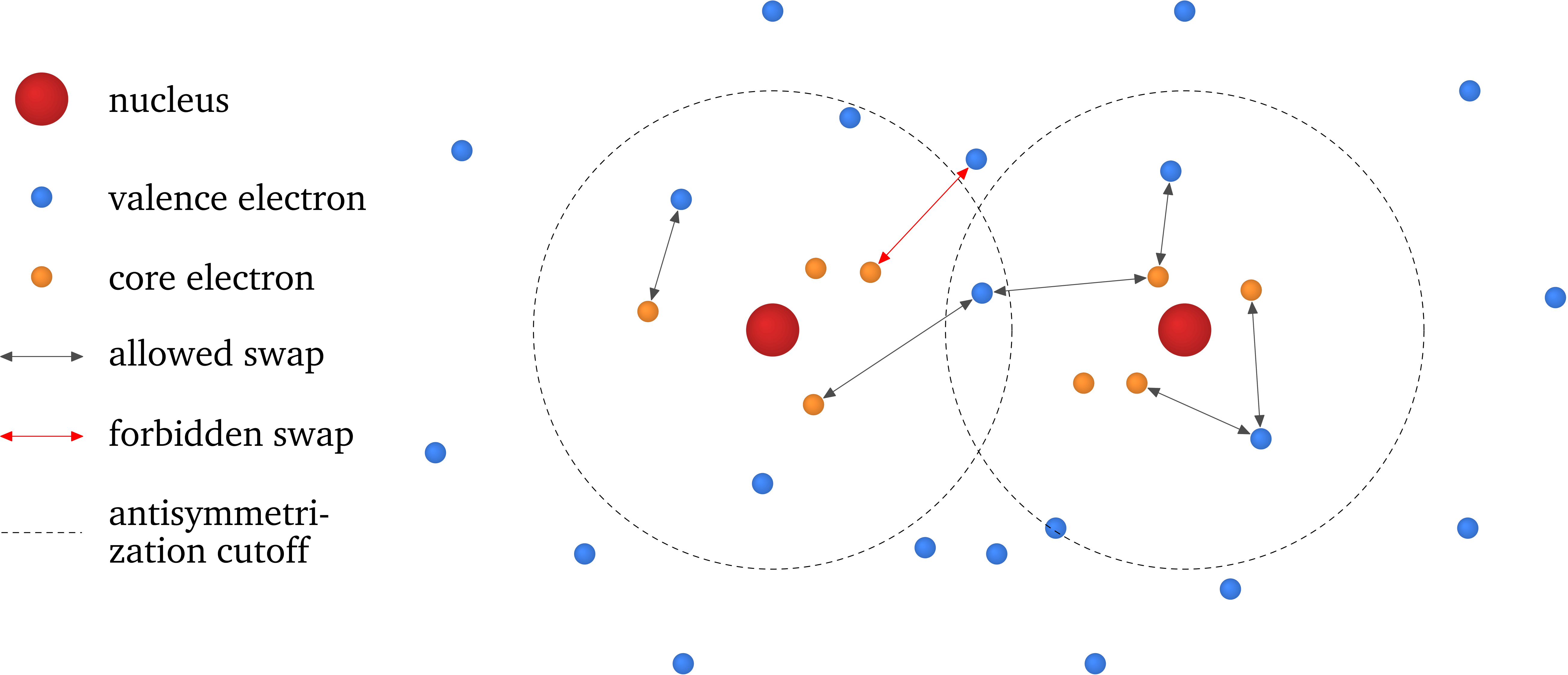}
    \caption{Possible approximation scheme of the antisymmetrizer would include labeling some electrons as core electrons based on their distance from the nucleus and only allowing a certain number of core-valence swaps within the antisymmetrization cutoff of each nucleus. Not all the forbidden/allowed swaps are shown for illustration for clarity.}
    \label{fig:antisymmetrizer_approximation}
\end{figure}

This approximation could be further extended to molecular systems by defining a spatial antisymmetrization cutoff around each nucleus and only allowing core-valence swaps within this cutoff. Because core-WF value for distant electrons is near-zero, this restriction would introduce negligible error. While this idea is illustrated schematically in figure \ref{fig:antisymmetrizer_approximation}, a full investigation is left for future work as it falls outside the scope of this study.

\subsection{Separating WF without an antisymmetrizer by electron labeling} \label{sec:without_anti}
An alternative way to separate the system into core and valence WFs could include integrating the separation directly into the orbital construction. This would include classifying electrons as core or valence based on, for instance, their distance from the nucleus, and setting the cross terms in the Slater determinant to zero, similar to the approach of \citet{scherbela_accurate_2025}. We can set the many-body orbitals to
\begin{equation} \label{eq:slater_separation}
    \phi_i(\br_j,\{\mr\}) = \begin{cases}
        \phi_i(\br_j,\{\mr_\text{C}\}) & \text{if } i,j \in \text{C}\\
        \phi_i(\br_j,\{\mr_\text{V}\}) & \text{if } i,j \in \text{V}\\
        0 & \text{otherwise}
    \end{cases}\,,
\end{equation}
where $\text{C}$ and $\text{V}$ denote sets of core and valence electrons respectively. This construction preserves the required equivariance condition
\begin{equation}
    \mathcal{P}_{jk} \phi_i(\br_j,\{\mr\}) =  \phi_i(\br_k,\{\mr\})\,.
\end{equation}
Substituting the definition \eqref{eq:slater_separation} into the Slater matrix, it can be rearranged to a block diagonal form, which in turn factorizes into a simple product of core and valence wave functions. Assuming a single Slater determinant for simplicity, we get
\begin{align*}
    \psi(\mr) &= \det \Phi(\mr) = \det \begin{pmatrix}
\phi_1(\br_1,\{\mr_\text{C}\}) & \cdots & 0 \\
\vdots & \ddots & \vdots \\
0 & \cdots & \phi_N(\br_N,\{\mr_\text{V}\})
\end{pmatrix}\\
    & = \det \Phi_\text{C}(\mr_\text{C}) \det \Phi_\text{V}(\mr_\text{V}) = \psi_\text{c}(\mr_\text{C})\psi_\text{v}(\mr_\text{V})\,,
\end{align*}
where we assumed the core electrons are ordered first. For other orderings, the result is unchanged up to a sign

This method completely eliminates the combinatorial cost associated with antisymmetrization. However, unlike the approximation discussed in section \ref{sec:speedup}, this approach introduces discontinuities when the distances of the outermost core electron and innermost valence electron become similar. These discontinuities lead to problematic estimates of the kinetic energy, as Monte Carlo walkers struggle in regions with sharp transitions and cannot properly account for the energy contribution caused by the discontinuity. A similar issue was encountered by \citet{richter-powell_sorting_2023}. For this reason, we did not explore this approach further. 
\section{Wave function ansatz adjustments}\label{app:WF_ansatz}

The architecture of our ansatz is based on the Psiformer ansatz\cite{von_glehn_self-attention_2022} implemented in {\sc DeepQMC} codebase\cite{schatzle_deepqmc_2023} that is based on JAX library\cite{jax2018github}. In this section, we outline the specific adjustments made to adapt the Psiformer architecture to our partitioning approach. We only discuss the adjustments and refer the reader to \citet{von_glehn_self-attention_2022} for a complete description of Psiformer itself.

\subsection{Exponential envelopes initialization} 
Most of the deep-learning ansätze (including Psiformer) employ envelopes -- simple scalar functions that multiply the many-body orbitals (i.e., the outputs of the graph neural network) before entering into the Slater determinant. These envelopes contain several trainable parameters and are essential for ensuring correct asymptotic behavior. Typically, envelopes are built as sums of exponential functions of type
\begin{align} \label{eq:standard_envelope}
    \varphi(r) = \eta \exp(-\zeta r)
\end{align}
where $r = \abs{\br_i - \boldsymbol{R}_I}$. The parameters $\eta$ and $\zeta$ are trainable and initialized with $\zeta = 1$, with separate sets of parameters for each nucleus.

In our approach, however, core-WF and valence-WF should have a qualitatively different behavior. The core-WF is expected to have high absolute values close to the nucleus and decay rapidly when moving away from it. On the other hand, the valence-WF is generally much wider and features a dip of its magnitude in the vicinity of the nucleus. See Figure \ref{fig:example_density} for illustration. To reflect these distinctions, we initialize the exponential parameters with $\zeta = 2$ for core-WFs and $\zeta = 0.5$ for the valence-WF. This initialization improves the energy convergence speed and numerical stability in practice. It also provides a convenient mechanism for designating which of the group wave functions represents the core and which represents the valence domain.

\subsection{Ensuring correct cusps}
In addition to ensuring correct asymptotic behavior at large distances, the exponential envelopes defined in \eqref{eq:standard_envelope} also introduce a cusp at $r=0$. Although the exact cusp condition is analytically known\cite{kato_eigenfunctions_1957}, it is not directly built-in, rather the ansätze typically learn the correct cusp without difficulty.

However, as the split-WF contains products of wave functions, this would lead to product of functions with a cusp at the same point. On top of that, the valence-WF should remain smooth at the nuclei. We therefore want to remove the cusp from valence-WF and only keep it in the core-WF for the particular nucleus. To achieve this, we modify the envelope function as follows
\begin{align} \label{eq:modified_envelope}
    \varphi_\text{cuspless}(r) = \eta \left(\exp(-\zeta r) - s \exp(-\zeta r / s) \right)
\end{align}
where $s \in (0,1)$ is a hyperparameter that we set to $s=0.1$ ensuring the envelope remains smooth and deviates from \eqref{eq:standard_envelope} only near the nucleus to avoid the cusp. For each nucleus we use exactly one group-WF with envelope given by \eqref{eq:standard_envelope} and the envelopes of other group-WFs are given by \eqref{eq:modified_envelope}. For instance, in the BeH molecule, the valence-WF uses a cuspless envelope at the Be nucleus and a regular envelope at the H nucleus, while the opposite applies to the core-WF.

\subsection{Explicit localization of core-WFs}
The core-WFs are expected to be localized around their associated nucleus without a strong dependence on the positions of other nuclei. Therefore, we exclude the electron-nucleus distance input features from the GNN input, except for those corresponding to the respective core. This ensures that the GNN of each core-WF explicitly receives information only from its own associated nucleus and no other nucleus. Nevertheless, it can still implicitly infer the presence of other nuclei through the electron coordinates, or rather from the electron density.


\subsection{Scaled-down Psiformer ansatz}
Ideally, the full Psiformer ansatz should be used as it is currently considered the most expressive type of ansatz. However, it is rather computationally expensive. While its expressivity is particularly beneficial for modeling larger molecular systems, it is not necessarily required for smaller systems. To keep the calculations tractable, we employed a scaled-down version of Psiformer for our experiments. We changed the following parameters of the Psiformer configuration:
\begin{itemize}
    \item number of determinants: $16 \rightarrow 4$,
    \item embedding dimension: $256 \rightarrow 64$,
    \item number of message passing GNN interactions: $4 \rightarrow 2$,
    \item attention heads: $4 \rightarrow 2$.
\end{itemize}
These simplifications reduced the number of training parameters for the LiH molecule from $1.6$ million to $52$ thousand and led to  an overall speedup of more than $3$-times (over $10$-fold for Be) for training on one NVIDIA RTX3090 GPU with only a minor loss in accuracy for these small systems.

\subsection{Hyperparameters of the experiments}

The adjusted Psiformer ansatz described in the previous section was employed in all the experiments in this work, and the training was performed using the default {\sc DeepQMC}\cite{deepqmc_2024} hyperparameters, unless otherwise noted. The relevant hyperparameters are summarized in table \ref{tab:experiment_hyperparemeters}.

\begin{table}[htb]
    \centering
    \begin{tabular}{cc}
        \toprule
         ansatz                     & adjusted Psiformer\\
         electron sample size       & 1\,000 \\
         pretraining iterations     & 0 \\
         training iterations\footnote{Only a half of the iterations ($5\!\cdot\!10^4$) were used for second-row atoms and only $2\!\cdot\!10^4$ for trainings with fixed core-WF.}    & $10^5$ \\
         learning rate at step $n$  & $\frac{0.05}{1 + (n / 10\,000)}$ \\
         evaluation iterations      & 5\,000 \\
         MC decorrelation steps     & 30 \\
         K-FAC decay                & 10 000 
         \\\bottomrule
    \end{tabular}
    \caption{Hyperparameters of the experiments}
    \label{tab:experiment_hyperparemeters}
\end{table}

We note that while generating Figure \ref{fig:energy_errors}, a few calculations faced numerical instabilities due to inconvenient number of core electrons. Restarting from the last checkpoint often resolved the issue. If this approach failed, we restarted the calculation from the beginning with a different seed. In rare instances where numerical instabilities persisted, we used {\tt AdamW}\cite{loshchilov_decoupled_2019} optimizer instead of {\tt K-FAC}\cite{botev_kfac-jax_2022} to mitigate this issue. For larger systems involving three group-WFs, we used {\tt AdamW}, as it led to more stable training.

In some experiments, we used the original non-modified Psiformer architecture with increased number of training steps and a batch size of 4\,000 instead of 1\,000. We refer to this configuration as \textit{accurate WF}, intended as a high-accuracy reference calculation.
\section{Electron density}\label{app:shape}


\subsection{Computation of one body density using KDE}
We work with spin-assigned wave function $\psi(\br_1,\dots,\br_N)$, where the first $N^\uparrow$ electrons are assigned spin up and the remaining $N^\downarrow$ are assigned spin down. The total electron density is then given by
\begin{align*}
    \rho(\br) &= N^\uparrow \int\abs{\psi(\br, \br_2, \dots,\br_N)}^2 \dd \br_2 \dots \dd \br_N\\
    &+N^\downarrow \int\abs{\psi(\br_1, \dots, \br_{N-1}, \br)}^2 \dd \br_1 \dots \dd \br_{N-1}\,. 
\end{align*}
To compute $\rho(\br)$, we first evaluate $\psi$ using several thousand evaluation iterations (i.e., without optimizing the WF parameters), storing the coordinates of the sampled electronic configurations. Using 1\,000 Monte Carlo walkers, we collect several million samples in total. 

For single-atom systems, we exploit spherical symmetry whenever applicable. We first compute the electron-nucleus distances for all electrons of all samples and apply one-dimensional Kernel Density Estimation\cite{silverman_bw_density_1986} (KDE) with Gaussian kernels to obtain a smooth function of the radial density profile. We employ the \texttt{stats.gaussian\_kde} function implemented in the \texttt{scipy} library\cite{2020SciPy-NMeth}.

In systems lacking spherical symmetry (these are the BeH and Li2 molecules in figures \ref{fig:BeH_different_geometries} and \ref{fig:Li2_c2c2v2}), we employ three-dimensional KDE and plot a one-dimensional slice of the resulting probability density function. Since estimating smooth densities in low-density regions requires more samples, these molecular plots appear slightly noisier compared to their atomic counterparts.



\subsection{Comparison of different core-valence splittings} \label{ssec:splitting_density_comparison}

\begin{figure}[tbh]
    \centering
    \includegraphics[width=0.8\linewidth]{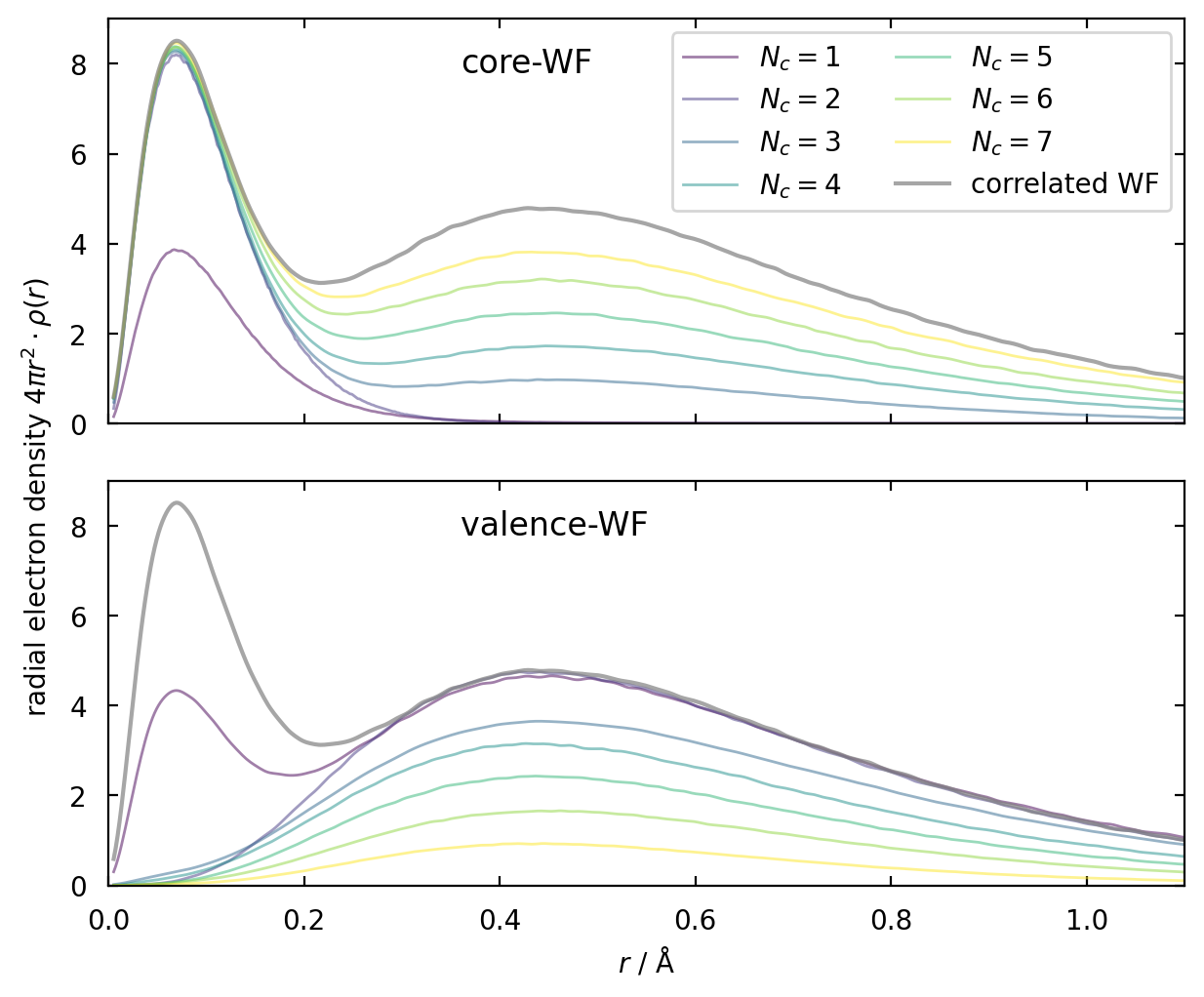}
    \caption{Comparison of electron densities of different numbers of core electrons in WF partitioning of oxygen atom. Clear separation is achieved for $N_c=2$ while other choices lead to significantly overlapping densities.}
    \label{fig:splitting_density_comparison}
\end{figure}

Figure \ref{fig:splitting_density_comparison} shows the comparison of densities of core- and valence-WF on an example system of oxygen atom with various numbers of core electrons $N_\text{c}$. When $N_\text{c} > 2$, the core-WF density leaks into the valence region, while its density in the core region remains nearly unchanged. $N_\text{c}=1$ is a special case, where the core-WF can only represent half of the core because it lacks the second electron. Only the splitting with $N_\text{c} = 2$ resulted in a distinct spatial separation between core and valence densities. In other cases, both densities remain mixed either in the core or valence domain. 

\begin{figure}[bth]
    \centering
    \includegraphics[width=0.8\linewidth]{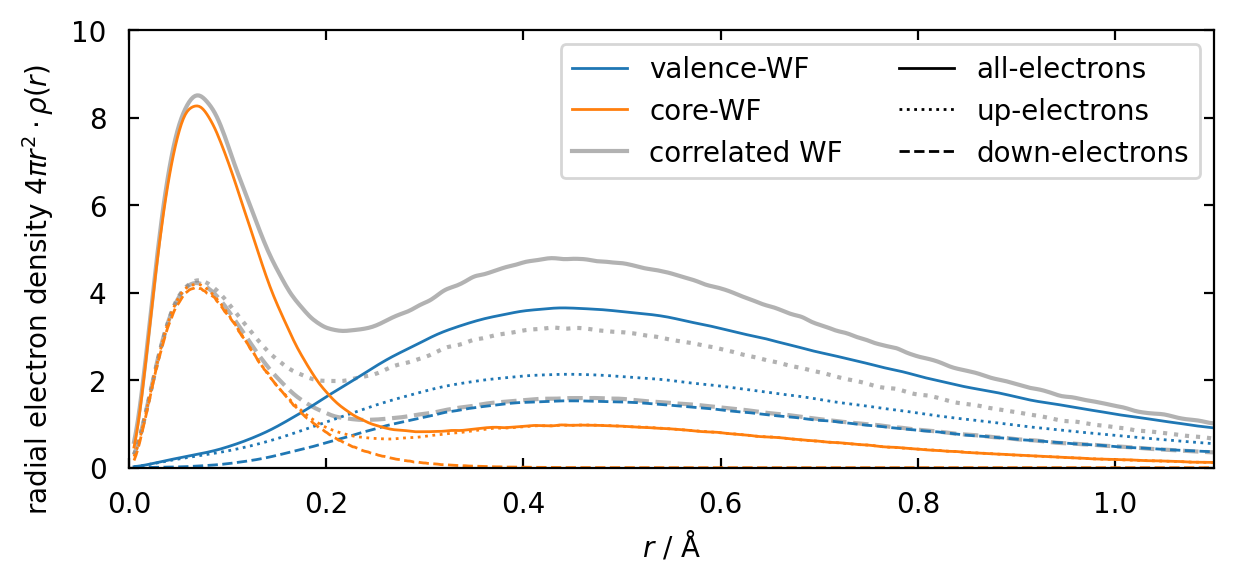}
    \caption{Decomposition of oxygen electron density into up and down electrons for 3 core + 5 valence electron partitioning. While the up-electrons (dotted lines) are mixed because some core up-electrons are present in the valence domain, the down-electrons (dashed lines) remain well separated between both domains.}
    \label{fig:splitting_up_down_densities_O}
\end{figure}

Since we work with spin-assigned WFs, we can direcly analyze the one-body densities of up and down electrons separately. Figure \ref{fig:splitting_up_down_densities_O} illustrates this for the oxygen atom using split-WF ansatz with three core and five valence electrons. While the total core and valence densities appear to overlap in the valence region, a different picture emerges when we examine only the spin-down component (there is one core down electron and two valence down electrons), where the core and valence densities are clearly separated. This indicates that the mixing originates from the up electrons. Specifically, our ansatz assigns two up electrons to the core by construction, although only one up electron should be classified as core. As a result, the extra spin up electron is pushed into the valence region, where it interacts indirectly with the valence electrons. However, the split-WF ansatz is unable to efficiently account for the correlations caused by these interactions, leading to suboptimal variational energy.

\section{Spherical symmetry of the Be wave function}\label{app:spherical_symm}

In this section, we investigate the symmetry properties of different WFs of the Beryllium atom. It has an interesting property because its exact ground state is a singlet state, which obeys spherical symmetry. Therefore, the wave function should remain unchanged under arbitrary rotations of the electronic configuration $\psi_\text{exact} (\mr) = \psi_\text{exact}(\mathrm{R}\mr)$. Surprisingly, the trained valence-WF does not accurately obey this spherical symmetry. We define a \textit{symmetry score} which measures the average variance of the wave function value when subjected to a rotation of samples
\begin{align} \label{eq:symmetry_score}
    \text{symmetry score} := \mathbb{E}_{\mr \sim \abs{\psi(\mr)}}\left[ \text{Var}_{\mr' \in \mathrm{R} \mr} \log \psi(\mr') \right] \,,
\end{align}
where $\mathrm{R}$ is a rotation matrix uniformly sampled\cite{shoemake_iii6_1992} from the SO(3) group.

\begin{figure}[bth]
    \centering
    \includegraphics[width=0.7\linewidth]{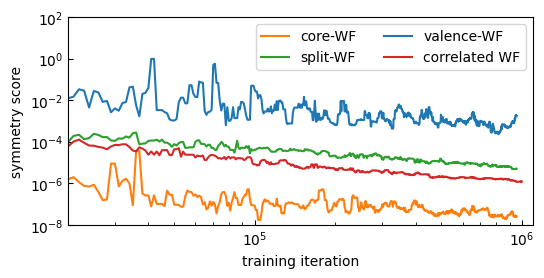}
    \caption{Spherical symmetry evolution during training of correlated ansatz (red) and split-WF (green). The symmetry of individual core and valence parts of the split-WF ansatz is shown in blue and orange respectively. A lower symmetry score indicates a higher degree of spherical symmetry in the wave function. The symmetry score is calculated using \eqref{eq:symmetry_score} and its median over the last 5 \% of iterations is plotted.}
    \label{fig:WF-split_training_sphericalSymmetry}
\end{figure}

The level of symmetry during training is shown in figure \ref{fig:WF-split_training_sphericalSymmetry}. We can see that the standard correlated calculation is slightly more symmetric than the split-WF ansatz. However, when we decompose the split-WF into individual core and valence parts, we can observe that core-WF is highly symmetric, whereas the valence-WF struggles to learn the spherical symmetry accurately. We note that while the energy has converged and does not exhibit any further observable improvement beyond the first 50 thousand training steps, the symmetry score continues to decrease (i.e., symmetry improves) as shown in figure \ref{fig:WF-split_training_sphericalSymmetry}.

These findings are relevant for applications such as ECP generation (details are given in Appendix \ref{app:fitting_bfd}), where the core-WF may be a promising candidate for fitting an ECP. If we wanted to use valence-WF, a significantly longer training period may be necessary to achieve a satisfactory level of symmetry in the valence-WF.

\section{Using split-WF to fit new ECPs} \label{app:fitting_bfd}

The split-WF method established a natural way for decomposing a many-body system into several electron groups. A similar problem is addressed in the process of generating the effective core potentials, which aim to separate core electrons from the calculations. In contrast to an ECP, the {split-WF} approach achieves this separation without relying on reference data or imposing artificial cutoff radii on the eigenfunctions. 

This naturally raises the question of whether one could exploit the full access to core-WF, valence-WF or both to construct a new ECP that replicates the effect that core-WF has on valence-WF. Such an ECP could enable efficient calculations on larger molecules. Furthermore, because the partitioning method can be applied to molecules, the resulting ECP could, in principle, take into account the molecular environment, adjusting its effect on the valence-WF in the presence of nearby atoms.

In this appendix, we introduce a pathway toward a new method for fitting an ECP based on the valence-WF. While this approach does not yet yield a fully accurate new ECP, it offers valuable insights into the properties of the valence-WF and conventional ECPs.

\subsection{Variance minimization method for ECP fitting}
The all electron molecular Hamiltonian in the Born--Oppenheimer approximation is given by
\begin{equation}\label{eq:hamil}
    \hat{H} = -\frac{1}{2} \sum_i^N \nabla_i^2 + \sum_{i < j}^{N} \frac{1}{\abs{\br_i - \br_j}} - \sum_{i,I}^{N,M} \frac{Z_I}{\abs{\br_i - \boldsymbol{R}_I}} \,.
\end{equation}
Traditional ECP methods replace the nuclear Coulombic attraction (last term) with an ECP operator denoted as $\hat{V}_\text{ecp}^\Theta$. Its parameters $\Theta$ are typically optimized to reproduce correct energies\cite{burkatzki_energy-consistent_2007, bennett_new_2017} or such that its orbitals match the all electron orbitals beyond a certain cutoff radius\cite{troyer_computational_2005, trail_correlated_2015, bachelet_pseudopotentials_1982}. The modified Schödinger equation for the wave function with effective core potential (referred to as the ecp-WF, denoted $\psi_\text{ecp}$) is written as
\begin{equation}
    \left(\hat{H}_\text{vlnc} + \hat{V}_\text{ecp}^\Theta \right) \ket{\psi_\text{ecp}} = E_\text{ecp} \ket{\psi_\text{ecp}}\,,
\end{equation}
where $\hat{H}_\text{vlnc}$ is an electronic Hamiltonian \eqref{eq:hamil} without the last Coulombic term and with $n<N$ electrons. In a conventional ECP calculation, both $\hat{H}_\text{vlnc}$ and $\hat{V}_\text{ecp}^\Theta$ are fixed operators, and the eigenstate $\psi_\text{ecp}$ is obtained by solving the above equation.

To fit a new ECP, we replace the ecp-WF with a trained valence-WF and seek an ECP operator that satisfies such modified Schrödinger equation. This involves first performing a reference calculation to optimize the split-WF, after which the resulting valence-WF is treated as reference data for fitting a new ECP. 
This setup leads to an inverse problem: given a target wave function, we aim to identify a quantum operator for which this wave function is the ground state.

We leverage the fact that the energy variance is zero if and only if the wave function is an exact eigenstate. We use this as the basis for our loss function. Specifically, we define the loss as
\begin{equation}\label{eq:variance_minimization_loss}
    \mathcal{L}({\Theta}) =\!\! \underset{\mr \sim \abs{\psi_\text{v}}^2}{\mathbb{E}} \bigg[ \underbrace{\frac{\braket{\mr}{\hat{H}_\text{vlnc} + \hat{V}_\text{ecp}^\Theta | \psi_\text{v}}}{\braket{\mr}{ \psi_\text{v}}}}_{\text{local energy}} - E^\text{EWM} \bigg]^2\,,
\end{equation}
where $\psi_\text{v}$ is the trained valence-WF. To avoid bias in the gradient estimates of $\mathcal{L}$, we replaced the mean energy with an exponentially weighted average of the mean energy over previous training steps $E_\text{EWM}$. This loss function enables us to optimize the parameters of $\hat{V}_\text{ecp}^\Theta$ term without the knowledge of ECP energy $E_\text{ecp}$. Although $\mathcal{L}({\Theta})=0$ holds not only for the ground state but for any eigenstate, we expect the loss function \eqref{eq:variance_minimization_loss} to be suitable, if the target wave function $\psi_\text{v}$ is well-behaved, because the electronic excited states would include a more complicated nodal structure.

\subsection{Form of ECP operator \texorpdfstring{$\hat{V}^\textnormal{ecp}_\Theta$}{Vecp}} \label{ssec:nnecp}

Here we discuss the \textit{ansatz} for the ECP operator, that is how we parametrize the operator with trainable parameters.
In most of the previous works, the ECP term was expressed in the following semi-local form
\begin{align}
    \hat{V}_\text{ecp}^\Theta &=\sum_I^M \sum_{i}^n \hat{V}_{Ii}^\Theta\,,\label{eq:ecp_form_separation}\\
    \hat{V}_{Ii}^\Theta &= V^{\Theta}_{\text{loc},I}(r_{Ii}) + \sum\limits_{\ell=0}^{\ell_\text{max}} V^{\Theta}_{\ell,I}(r_{Ii}) \hat{P}^{i}_{\ell,I}\,, \label{eq:ecp_form}
\end{align}
where $\hat{P}^{i}_{\ell,I}$ is an angular momentum projector of $i$-th electron onto the subspace of angular quantum number $l$ with respect to the $I$-th nucleus. This form allows the ECP operator to act differently on electrons with different angular momenta thus mimicking the effect of different core subshells. The scalar functions $V^{\Theta}_{\text{loc},I}(r)$ and $V^{\Theta}_{\ell,I}(r)$ are conventionally represented by a sum of a few Gaussians whose parameters (usually dozens in total) are obtained via various fitting strategies.

We keep the form of \eqref{eq:ecp_form} and introduce much more flexible ECP ansatz representing functions $V^{\Theta}_{\text{loc},I}(r)$ and $V^{\Theta}_{\ell,I}(r)$ by an MLP with two hidden layers. The MLP input is the distance $r$, expanded using Gaussian basis features. It produces $1+\ell_\text{max}$ outputs, which are interpreted as functions $V^{\Theta}_{\text{loc},I}(r)$ and $V^{\Theta}_{\ell,I}(r)$. At large distance $r$, we enforce the correct asymptotic behavior for $r\rightarrow\infty$ by transitioning these functions to effective Coulomb attraction or Gaussian decay. We also increase $\ell_\text{max}$ to enlarge the operator space spanned by our ECP ansatz. In such a case, however, it is necessary to include a small penalty for larger $l$ to favor ECP operators with lower angular momentum.

While conventional ECPs typically use only dozens, our ECP ansatz with tens of thousands of parameters offers more flexibility. However, it is still limited by the conventional ECP form \eqref{eq:ecp_form} adapted for individual treatment of different angular momenta. We kept the form \eqref{eq:ecp_form} because it allows for efficient computation of angular momenta integrals via quadrature rules\cite{li_fermionic_2022}. Implementing a general non-diagonal quantum operator $V_\text{ecp}^\Theta (\mr; \mr')$ would require computing $6N$-dimensional integrals, which would be too expensive even in a Monte Carlo setting.


\subsection{Fitting ECP term from ecp-WF}

To test the above-described variance minimization method and the validity of our inverse problem, we first carry out the following experiment. A standard deep QMC calculation on a Be atom with a conventional ECP from the literature is performed to obtain a $\psi_\text{ecp}$. Once we have an accurate $\psi_\text{ecp}$, we \textit{forget} the ECP operator and optimize a new $\hat{V}_\text{ecp}^\Theta$ operator from scratch that is represented by our neural network ansatz defined in section \ref{ssec:nnecp} using the loss \eqref{eq:variance_minimization_loss}.

\begin{figure}[htb]
    \centering
    \includegraphics[width=0.85\linewidth]{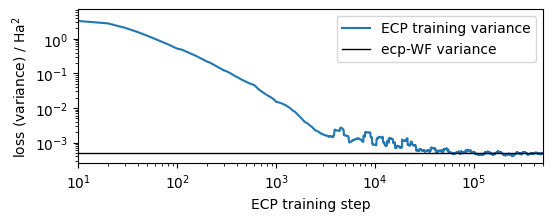}
    \caption{Evolution of energy variance given by \eqref{eq:variance_minimization_loss} during ECP training. A variance of the conventional ecp-WF (calculated with ECP term from literature) providing the optimal learning target is shown as black line.}
    \label{fig:s2_energy_convergence}
\end{figure}

\begin{figure}[htb]
    \centering
    \includegraphics[width=0.85\linewidth]{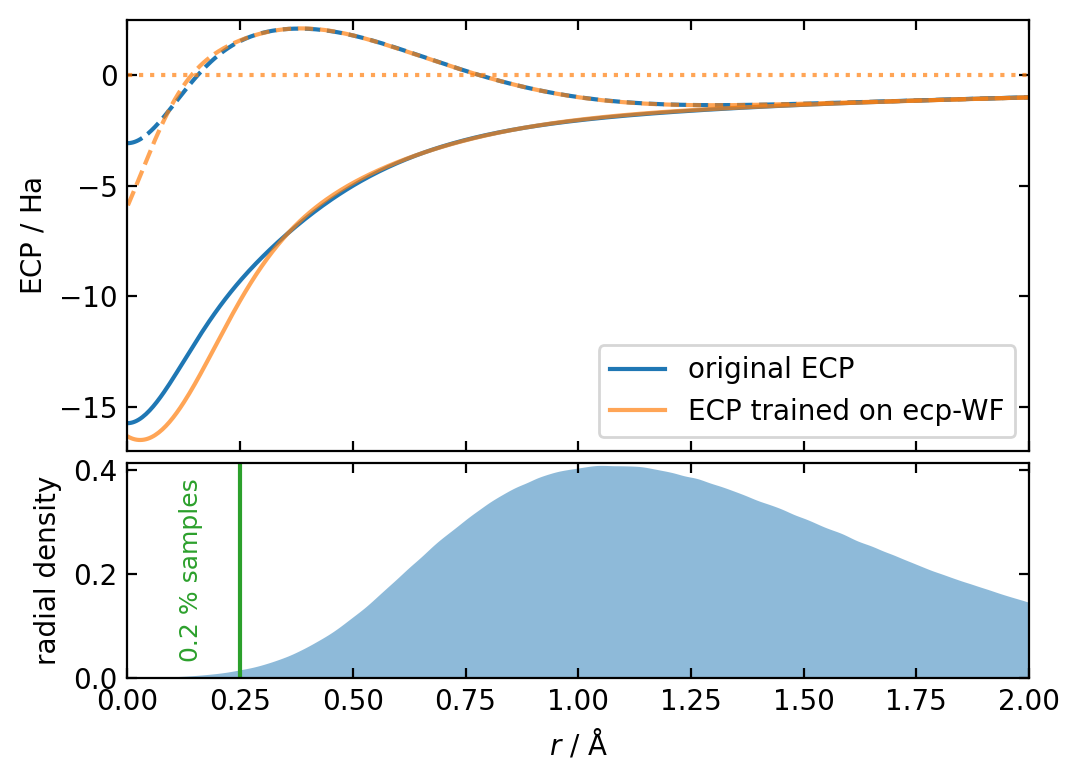}
    \caption{Shapes of conventional ECP functions from the literature\cite{bennett_new_2017} (orange lines) compared with the ECP functions recovered by our neural network (blue lines). The network has no direct access to the correct ECP functions, it was trained on ecp-WF that only contains an indirect information about ECP. The solid lines show $V_\text{loc}^\Theta(r)$ function and dashed lines show the zeroth angular momentum channel $V_\text{loc}^\Theta(r) + V_0^\Theta(r)$. The dotted line depicts $V_1^\Theta(r)$ of our trained ECP, which is close to zero, which is a correct behavior as the conventional ECP does not have an $l=1$ channel. Radial distribution of electron samples is shown in the bottom panel.}
    \label{fig:bfd_trained_shapes}
\end{figure}

Using the AdamW optimizer, we were able to minimize the loss to reach the value of the energy variance that matches the final energy variance of $\psi_\text{ecp}$ obtained in the reference calculation with the conventional ECP operator. This hints that a good optimum has been reached, see Figure \ref{fig:s2_energy_convergence}. Figure \ref{fig:bfd_trained_shapes} compares the conventional ECP functions $V^{\Theta}_{\text{loc}}(r)$ and $V^{\Theta}_{\ell}(r)$ used to train $\psi_\text{ecp}$ and the functions learned by our neural network. The functions are in close agreement except in the region close to the nucleus. This is explained by the fact that there are very few electron samples in the vicinity of the nucleus as the valence electrons are repelled by the effective potential, making the ECP shape in that region less relevant, and therefore not perfectly reproduced by our ansatz. 
Nevertheless, we showed that our variance minimization method can recover the ECP term just from the ecp-WF, confirming the validity of our approach.

\subsection{Outlook for ECP fitting}

Finally, we investigated whether this method can be used to find a new ECP term that would correspond to valence-WF obtained from a split-WF reference calculation. This would open a new way of ECP generation. However, despite considerable efforts, we were unable to minimize the energy variance \eqref{eq:variance_minimization_loss} to a satisfactory level to provide us with an operator whose ground state is $\psi_\text{v}$. Two key factors may be responsible for this problem. 


Firstly, the problem might stem from the fact that the individual group-WFs are not unique, as discussed in \ref{sec:non_uniquness_problem}. It is conceivable that the ambiguity of core-valence splitting could by resolved by adding an appropriate penalty term, that would guide the valence-WF to a more favorable form during the split-WF optimization.

Secondly, although our ECP ansatz $\hat{V}_\text{ecp}^\Theta$ is much more flexible than traditional ECP ansätze, it can still only represent a very specific class of semi-local quantum operators that can be decomposed into individual angular momentum channels via eq. \eqref{eq:ecp_form_separation}.
A general non-local quantum operator cannot be decomposed into such a form. Lifting the restriction \eqref{eq:ecp_form_separation} and employing a general operator $V_{\text{ecp}}^\Theta (\mr;\mr')$ would be too costly.

\end{document}